\documentclass[aps,pra,showpacs,twoside,twocolumn,longbibliography,10pt]{revtex4-2}
\usepackage[colorlinks=true, citecolor=red, urlcolor=blue ]{hyperref}
\usepackage{epsfig,newlfont,amssymb,amsfonts,amsmath,bm,palatino,mathtools,amsthm,braket,times,soul,enumitem}
\usepackage[normalem]{ulem}
\newcommand{\stkout}[1]{\ifmmode\text{\sout{\ensuremath{#1}}}\else\sout{#1}\fi}
\usepackage[english]{babel}
\usepackage[utf8]{inputenc}
\usepackage{xcolor}
\usepackage{graphics}

\usepackage{verbatim}
\usepackage{bbm}
\usepackage{wrapfig}
\usepackage{pifont}

\newcommand{\cmark}{\ding{51}} % ✓
\newcommand{\xmark}{\ding{55}}

\usepackage{hyphenat}

\usepackage{orcidlink}

\newlength\figureheight 
\newlength\figurewidth

\begin{document}

\title{Pulse Engineering of Quantum Many-Body Dynamics: \\ Emergent Scar States, Entanglement, and Nonstabilizerness}

\author{Prasant Mallik}
\author{Arkaprava Sil}
\author{Sudipto Singha Roy}
\affiliation{Department of Physics, IIT (ISM) Dhanbad, India}

\date{\today}
\begin{abstract}
Understanding and coherently controlling the properties of interacting quantum many-body systems is a central challenge in non-equilibrium quantum physics. While in the past decades, a wide range of many-body Hamiltonians have been introduced to study quantum chaos, atypical eigenstates, and quantum resources, systematically engineering and continuously tuning these properties within a single microscopic model remains largely unexplored. Here, we employ a pulse engineering scheme to construct an effective Hamiltonian that continuously interpolates between a chaotic Heisenberg (XXX) chain with a local impurity and the ZX Hamiltonian. Along this interpolation, we identify several families of analytically tractable atypical eigenstates embedded in the excited-state spectrum with distinct entanglement and nonstabilizerness properties. In the XXX limit, these states exhibit exact plateaus in both entanglement and stabilizer R\'enyi entropy and correspond to coherent superpositions of long-range valence-bond solid (VBS) states. As the pulse strength increases, the effective Hamiltonians exhibit a hierarchy of new set of approximate entanglement plateaus in the low-energy spectrum. Interestingly, in the fully pulse-engineered ZX limit, we uncover a distinct pair of long-range entangled stabilizer eigenstates, corresponding to rainbow scar states. We further show that pulse engineering preserves the distinct chaotic and non-chaotic regimes of the original model, while that is largely absent in the dynamical generation of entanglement and nonstabilizerness. The pulse-engineered models generate nearly identical quantum resources in both regimes, revealing a partial decoupling between quantum chaos and quantum-resource generation. Our results establish pulse engineering as a versatile framework for generating many-body Hamiltonians with structured eigenstates and tunable quantum resources.
\end{abstract}

\maketitle

\section{Introduction}

The ability to coherently control and manipulate the dynamics of interacting quantum systems lies at the heart of modern quantum science and technology. In recent years, periodic driving and pulse-based control techniques~\cite{Haeberlen_1968,NMR_2005,Alessio_2013,Goldman_2014,bukov_2015, poudel_2015,Oka_2019,RudnerM_2020,Sen_2021} have emerged as powerful tools to engineer exotic quantum dynamics. Such protocols play a central role in diverse areas of quantum technology, ranging from decoherence suppression to quantum simulation~\cite{Lorenza_1999,Khodjasteh_2005,Uhring_2007,Khodjasteh_2009,du_2009,Sutar_2016}, and in the unveiling of exotic non-equilibrium many-body dynamics~\cite{choi_2020,Sahay_2021,Kranzl_2023,Nandy_2024}. For instance, Floquet engineering~\cite{Alessio_2013, bukov_2015,Oka_2019,RudnerM_2020, Sen_2021} provides a versatile route for modifying effective Hamiltonians and tailoring interaction structures through periodic driving, offering a powerful approach to engineer quantum many-body dynamics. 
A central theoretical framework underlying these approaches is the average Hamiltonian theory, originally developed in the context of nuclear magnetic resonance~\cite{Haeberlen_1968,brinkmann_2016}. In case of pulse driven approaches,  applying carefully designed pulse sequences~\cite{biercuk_2009_error,Khodjasteh_2010,West_f_2010,Wu_2023,Zhao_2025,Peterson_2020}, unwanted interactions and disorder effects can be selectively suppressed while desired couplings are enhanced~\cite{Fraval_2005,Frydrych_2014,Frydrych_2014,Louzon_2025,Brown_2025}. Beyond coherence preservation, these methods have enabled the realization of engineered many-body phenomena, including dynamical phase transitions~\cite{Jafari_2021,Naji_2022,Jafari_2022,Diptarka_2026}, discrete time crystals~\cite{zhang_2017,Randall_2021,Zaletel_2023}, etc. This has helped modern experimental platforms, including ultracold atoms~\cite{Weitenberg_2021}, superconducting qubits~\cite{Nguyen_2024}, Rydberg arrays~\cite{Labuhn_2016,Bluvstein_2021}, and trapped ions~\cite{Morong_2023}, access dynamical regimes that were once
purely theoretical. 
\begin{figure*}[t]
    \centering
    \includegraphics[width=0.9\linewidth]{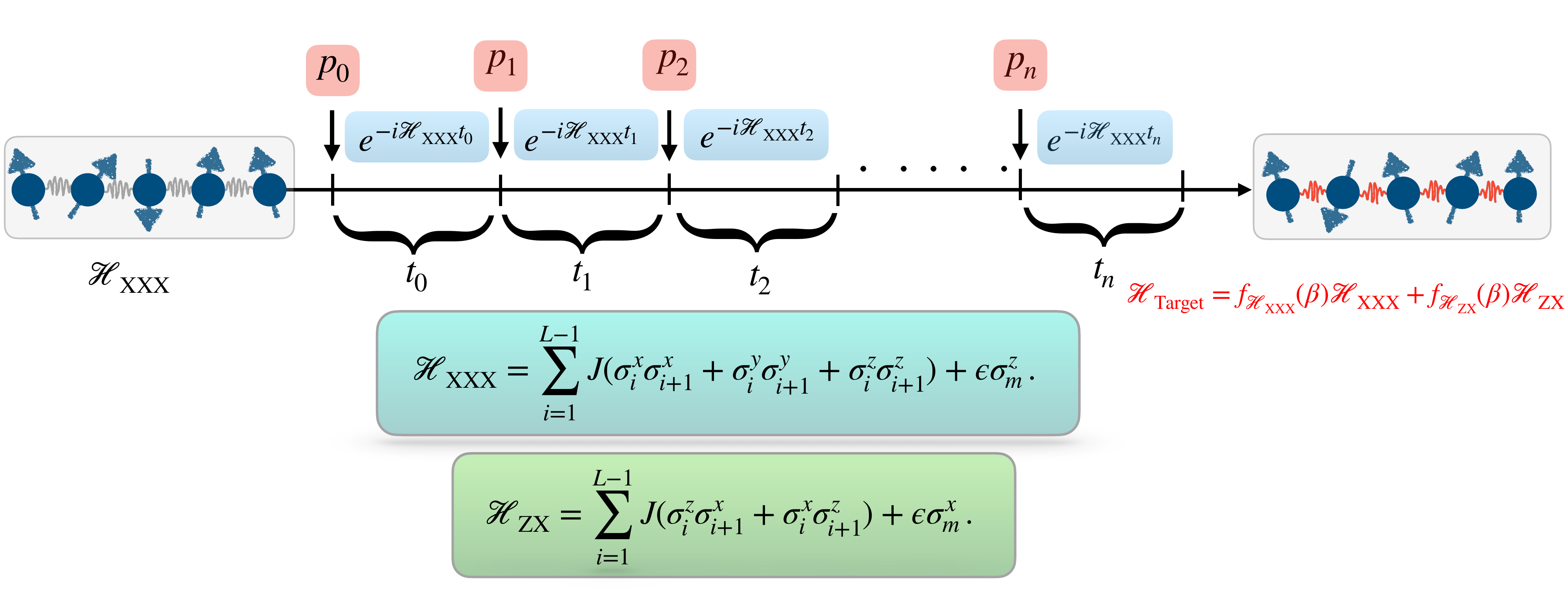}
    \caption{Schematic illustration of the pulse-driven spin chain and the effective Hamiltonian generated through the pulse-based Hamiltonian engineering protocol.}
    \label{fig:schematic}
\end{figure*}

Despite these advances, %, most previous studies on pulse engineering have focused mainly on controlling effective interactions~\cite{Frydrych_2014,Brown_2025}, noise suppression~\cite{Louzon_2025}, reducing decoherence~\cite{Lorenza_1999,Fraval_2005,West_2010}, or improving gate fidelities~\cite{Peterson_2020,biercuk_2009_error,Khodjasteh_2010,West_f_2010,Wu_2023,Zhao_2025}. \textcolor{blue}{seems repeat of senences}. 
 much less is understood about how coherent control influences quantum complexity and the structure of many-body eigenstates. Traditionally, these questions are investigated by tuning the parameters of microscopically different many-body Hamiltonians and analyzing the resulting changes in its spectral and dynamical properties. Such studies have established that, in generic  non-integrable systems, quantum chaos is often accompanied by thermal eigenstate behavior, rapid entanglement growth~\cite{Kim_2013,Nahum2017}, enhanced nonstabilizerness (magic)~\cite{Kanato_2022,Dowling_2025,Oliviero_2025}, etc. At the same time, there are notable exceptions to this picture. For example, atypical non-thermal eigenstates, namely,  quantum many-body scars~\cite{Turner_2018,Moudgalya_2022}, can coexist with otherwise chaotic spectra while exhibiting anomalous entanglement  properties~\cite{Langlett_2022,Kerschbaumer_2025}. Another class of atypical eigenstates emerges as the $\eta$-pairing states of the Hubbard model~\cite{Moudgalya_2020,Daniel_2020}, which remain non-thermal due to an underlying pseudospin $SU(2)$ symmetry.  These findings suggest that the relationship between quantum chaos, quantum resources, and eigenstate structure is richer than commonly anticipated.\\

In this regard,  pulse engineering offers a complementary route to investigate this interplay. Rather than modifying the microscopic Hamiltonian itself, it continuously deforms the effective Hamiltonian while leaving the underlying physical system unchanged. This provides a unique platform to track how eigenstates and quantum resources evolve under controlled Hamiltonian deformations, naturally leading to two key questions. First, how do structured many-body eigenstates evolve under such deformations? Second, how does pulse engineering reshape the connection between spectral chaos and quantum resource generation, particularly entanglement and nonstabilizerness?

To address these questions, we consider a chaotic spin-$1/2$ XXX chain with a single site  impurity~\cite{Rabinovici_2022} and construct pulse-engineered effective Hamiltonians using sequences of Clifford operations within the framework of average Hamiltonian theory. Besides their central role in quantum information processing and fault-tolerant quantum computation~\cite{NielsenChuang2010}, Clifford operations preserve the Pauli structure of the Hamiltonian while remaining experimentally accessible, thereby providing a natural route to analytically tractable Hamiltonian deformations. The resulting effective Hamiltonians continuously interpolate between the original chaotic XXX model and ZX Hamiltonian containing mixed $(\sigma_i^x\sigma_{i+1}^z+\sigma_i^z\sigma_{i+1}^x)$ interactions. This interpolation enables us to systematically track how spectral properties, quantum complexity, and eigenstate structure evolve under coherent control.

At the undeformed XXX limit, we uncover a family of $L-1$ analytically tractable atypical eigenstates for system size $L$ (where $L$ is odd) embedded within the chaotic spectrum. Using an effective Hamiltonian description, we derive exact analytical expressions for these states and show that they correspond to coherent superpositions of long-range valence-bond-solid configurations. These states give rise to exact plateaus in both the half-chain entanglement entropy and the stabilizer Rényi entropy. While the half-chain entanglement entropy of these states remain exactly equal to unity  for all system sizes, the corresponding stabilizer Rényi entropy increases systematically with system size. Beyond these exact plateaus, the static analysis also reveals a hierarchy of approximate entanglement plateaus in different excitation sectors~\cite{Alba_2009}. We further investigate the dynamical generation of quantum resources and find that the more chaotic regimes generally produce stronger entanglement and nonstabilizerness, consistent with the conventional picture of chaotic many-body dynamics.

As the pulse-induced deformation increases, these structured eigenstates evolve in a non-trivial manner. The exact high-energy plateau gradually disappears and only approximate plateaus survive in the low-energy sector, indicating enhanced thermalization.  At the same time, we find that the conventional correspondence between spectral chaos and quantum resource generation becomes progressively weakened. Hamiltonians with markedly different level statistics exhibit nearly identical entanglement and nonstabilizerness generation, pointing to a partial decoupling between spectral chaos and quantum complexity.  In the fully pulse-engineered limit, the effective ZX Hamiltonian hosts a qualitatively different family of structured eigenstates. These are long-range entangled stabilizer eigenstates embedded near the middle of the spectrum, commonly known as  the rainbow scar states reported previously~\cite{Langlett_2022}.

Taken together, our results demonstrate that pulse engineering provides a controlled route for continuously reshaping both many-body eigenstates and quantum complexity within a single microscopic model. Beyond modifying effective interactions, it enables one to drive the system between distinct classes of non-thermal eigenstates, namely, from analytically tractable VBS-like atypical states in the chaotic XXX chain to the long-range entangled rainbow scar states of the pulse-engineered ZX Hamiltonian, while also revealing how  entanglement, and nonstabilizerness evolve throughout the interpolation.  These findings establish coherent control not only as a versatile tool for Hamiltonian engineering but also as a powerful framework for creating and exploring unconventional many-body quantum states and quantum resources in near-term quantum devices.

The article is organized as follows. In Sec.~\ref{sec:model}, we introduce the pulse-engineered models and derive the corresponding effective Hamiltonian using the average Hamiltonian theory. Sec.~\ref{sec:symmetry} discusses the symmetry modifications induced by pulse engineering that is followed by the chaotic property analysis of the modified Hamiltonians in Sec.~\ref{sec:chaos}. We present analytically tractable structured eigenstates emerging in the pulse-engineered Hamiltonians in  Sec.~\ref{sec:static}.  Thereafter, the investigation of the interplay between spectral chaos, entanglement generation, and magic under the resulting dynamics is presented  in Sec.~\ref{sec:dynamics}. Finally, in Sec.~\ref{sec:discussion}, we summarize our main results and discuss possible future directions.

\section{Model and formalism}
\label{sec:model}
In this section, we introduce the interacting many-body Hamiltonian considered in this work and outline the pulse-engineering protocol used to construct the corresponding effective dynamics. Here, we consider a one-dimensional spin-$1/2$  chain described by the Hamiltonian
\begin{equation}
\mathcal{H}_{\mathrm{XXX}} =
\sum_{i=1}^{L-1}
J \left[\sigma^x_i \sigma^x_{i+1}+\sigma^y_i\sigma^y_{i+1}+\sigma^z_i\sigma^z_{i+1}
\right]
+
\varepsilon \sigma^z_m,
\end{equation}
where $L$ is taken to be odd, and  $m=(L+1)/2$ denotes the middle lattice site, $\{\sigma_i^x,\sigma_i^y,\sigma_i^z\}$ are the Pauli operators acting on site $i$, and $\varepsilon$ represents the strength of a local impurity field applied along the $z$-direction at the middle site.  It is already known that, for suitable parameter regimes, the model exhibits strong signatures of many-body quantum chaos~\cite{Rabinovici_2022}.   Throughout the numerical analysis of this work, we set $J=0.6$, which allows the ratio $J/\epsilon$ to span both the exchange-dominated ($J/\epsilon>1$) and impurity-dominated ($J/\epsilon<1$) regimes while keeping the impurity strength within a moderate parameter range. However, the analysis is not restricted to this particular choice of $J$ and remains valid for any moderate  values of $J/\varepsilon$.

To engineer modified interaction structures, we employ a pulse-based modulation protocol and analyze the resulting dynamics within the framework of average Hamiltonian theory. For completeness, we first briefly outline the general scheme and then specify the particular choices of pulses considered in this work.  For a sequence of pulse operations $p_0,p_1, \dots,p_n$ interspersed with evolution under the Hamiltonian $\mathcal{H}$ for durations $t_0,t_1, \dots,t_n$, the evolution operator over one driving cycle is given by
\begin{equation}
U_T
=
 e^{-i\mathcal{H} t_{n}}p_n
\cdots
p_1 e^{-i\mathcal{H} t_0}p_0.
\end{equation}
Here $T=\sum_k t_k$ is the duration of one driving cycle. Let us introduce the operators $g_n=p_np_{n-1}\dots p_0$, $p_n=g_ng^\dagger_{n-1}$. 
The evolution operator thus becomes
{
\begin{eqnarray}
 U_T
&=&
(p_n \cdots p_0) \Big[
(p_n \cdots p_0)^\dagger e^{-i\mathcal{H} t_n} (p_n \cdots p_0)\Big] \nonumber 
\\
&&\Big[(p_{n-1} \cdots p_0)^\dagger e^{-i\mathcal{H} t_{n-1}} (p_{n-1}\cdots p_0)\Big]\nonumber\\&&  \dots  \Big[(p_1p_0)^\dagger e^{-i\mathcal{H} t_{1}} (p_1p_0)\Big]
\Big[p_0^\dagger e^{-i\mathcal{H} t_0}p_0\Big] \nonumber \\
&=&g_n \Big[
 e^{-i (g_n^\dagger \mathcal{H} g_n)t_n}\Big] \Big[ e^{-i(g_{n-1}^\dagger \mathcal{H} g_{n-1}) t_{n-1}}\Big]  \dots \nonumber \\&&  \Big[ e^{-i(g_1^\dagger \mathcal{H} g_1) t_{1}}\Big]
\Big[ e^{-i (g_0^\dagger\mathcal{H}g_0) t_0}\Big] \nonumber \\
&=&g_ne^{-i\sum_k \mathcal{H}_{k}t_{k}}=e^{-i\sum_k \mathcal{H}_{k}t_{k}},
\label{eqn:evolution_exact}
\end{eqnarray}
}
where we have considered $g_n = I^{\otimes L}$ and $\mathcal{H}_{k} =g_k^\dagger \mathcal{H} g_k$. 

The average Hamiltonian theory implies that the evolution operator over one driving cycle can be written as
\begin{equation}
U_T=e^{-i\mathcal{H}_{\mathrm{eff}}T}, \ \mathrm{or}, \  \mathcal{H}_{\mathrm{eff}}=\frac{i}{T}\log(U_T).
\label{eqn:H_eff}
\end{equation}
where the effective Hamiltonian is given by the Magnus expansion~\cite{Haeberlen_1968,brinkmann_2016},
\begin{align}
\mathcal{H}_{\mathrm{eff}}
&=
\mathcal{H}_{\mathrm{eff}}^{(0)}
+\mathcal{H}_{\mathrm{eff}}^{(1)}
+\mathcal{H}_{\mathrm{eff}}^{(2)}
+\cdots \nonumber\\
&=
\frac{1}{T}\sum_k \mathcal{H}_k t_k
-\frac{i}{2T}\sum_{k>l}
[\mathcal{H}_k,\mathcal{H}_l]\,t_kt_l
+\mathcal{O}(T^2).
\label{eq:Magnus}
\end{align}
The leading-order (zeroth-order Magnus) term is therefore
\begin{equation}
\mathcal{H}_{\mathrm{Target}}=\mathcal{H}_{\mathrm{eff}}^{(0)}
=
\frac{1}{T}\sum_k
\mathcal{H}_k t_k
=
\frac{1}{T}\sum_k
g_k^\dagger \mathcal{H} g_k\, t_k.
\end{equation}

%In our protocol, the pulse operators {\color{red}act only on even lattice sites} and are chosen as
%\begin{align}
%p_0 &= I_1 \otimes H_2 \otimes I_3 \otimes H_4 \otimes \cdots , \nonumber\\
%p_1 &= I_1 \otimes (\sigma^yH)_2 \otimes I_3 \otimes (\sigma^yH)_4 \otimes \cdots , \nonumber\\
%%p_2 &= I_1 \otimes(\sigma^y)_2 \otimes I_3 \otimes(\sigma^y)_4 \otimes \cdots  
%\end{align}
%where $H_i$,  and $I_i$ denote Hadamard,  and identity operators acting on site $i$, respectively, and other operators have already been defined. The corresponding choices of the  $g_i$ operators are given by 
%\begin{align}
%g_0 &= I_1 \otimes H_2 \otimes I_3 \otimes H_4 \otimes \cdots ,\nonumber\\
%g_1 &= I_1 \otimes \sigma^y_2 \otimes I_3 \otimes \sigma^y_4 \otimes \cdots ,\nonumber \\
%g_2 &= I^{\otimes N}.
%\end{align}
In our protocol, we use three pulses $(p_0, p_1, p_2)$. Since we consider only an odd number of lattice sites $L$, two distinct cases arise depending on whether the middle site $m=(L+1)/2$ is even or odd. 

\noindent\textit{Case I: $m$ even.} When the middle site is even (e.g., $m=2,4,6,\cdots$), the pulse operators act exclusively on the even lattice sites, while the odd sites remain unchanged. The pulse sequence is given by
\begin{align}
p_0 &= I_1\otimes H_2\otimes I_3\otimes H_4\otimes\cdots, \nonumber\\
p_1 &= I_1\otimes(\sigma^{y}H)_2\otimes I_3\otimes(\sigma^{y}H)_4\otimes\cdots, \nonumber\\
p_2 &= I_1\otimes(\sigma^{y})_2\otimes I_3\otimes(\sigma^{y})_4\otimes\cdots,
\label{eq:pulses1}
\end{align}
where $H_i$,  and $I_i$ denote Hadamard,  and identity operators acting on site $i$, respectively, and other operators have already been defined. In  this case, corresponding $g_k$ operators are given  by 
\begin{align}
g_0 &= I_1\otimes H_2\otimes I_3\otimes H_4\otimes\cdots, \nonumber\\
g_1 &= I_1\otimes(\sigma^{y})_2\otimes I_3\otimes(\sigma^{y})_4\otimes\cdots, \nonumber\\
g_2 &= I^{\otimes L}.
\label{eq:g1}
\end{align}

\noindent\textit{Case II: $m$ odd.} When the middle site is odd (e.g, $m=3,5,7,\cdots$), the protocol is identical to Case I, except that the pulses act exclusively on the odd lattice sites, while the even sites remain unchanged. The corresponding pulse sequence is
\begin{align}
p_0 &= H_1\otimes I_2\otimes H_3\otimes I_4\otimes\cdots, \nonumber\\
p_1 &= (\sigma^{y}H)_1\otimes I_2\otimes(\sigma^{y}H)_3\otimes I_4\otimes\cdots, \nonumber\\
p_2 &= (\sigma^{y})_1\otimes I_2\otimes(\sigma^{y})_3\otimes I_4\otimes\cdots,
\label{eq:pulses2}
\end{align}
with the corresponding $g_k$ operators
\begin{align}
g_0 &= H_1\otimes I_2\otimes H_3\otimes I_4\otimes\cdots, \nonumber\\
g_1 &= (\sigma^{y})_1\otimes I_2\otimes(\sigma^{y})_3\otimes I_4\otimes\cdots, \nonumber\\
g_2 &= I^{\otimes L}.
\label{eq:g2}
\end{align}

Under this pulse sequence and for a particular choices of the $t_i$'s that we specify below,  the original Hamiltonian is transformed into the effective form
\begin{equation}
\mathcal{H}_{\mathrm{Target}}
=
f_{\mathcal{H}_{\mathrm{XXX}}}(\beta)\mathcal{H}_{\mathrm{XXX}}
+
 f_{\mathcal{H}_{\mathrm{ZX}}}(\beta) \mathcal{H}_{\mathrm{ZX}} ,
\label{eqn:modulated}
\end{equation}
where 
\begin{equation}
\mathcal{H}_{\mathrm{ZX}} =
\sum_{i=1}^{L-1}
J\left(
\sigma^z_i \sigma^x_{i+1}
+
\sigma^x_i \sigma^z_{i+1}
\right)
+
\varepsilon \sigma^x_m, 
\label{eq:H2_general}
\end{equation}
\noindent and $f_{\mathcal{H}_{\mathrm{XXX}}}(\beta)=(1-\beta)$, $f_{\mathcal{H}_{\mathrm{ZX}}}(\beta)=2\beta$. Here, $\beta=t_1/t_2$  is a dimensionless parameter that controls the relative durations of the pulse intervals within a pulse cycle of duration $T=t_0+t_1+t_2$. The pulse intervals are parameterized as
\begin{equation}
t_0=\frac{2\beta T}{1+3\beta},\qquad
t_1=\frac{\beta T}{1+3\beta},\qquad
t_2=\frac{T}{1+3\beta}.
\end{equation}

Here, we choose $\beta \in[0,1]$. Hence, when $\beta=0$,  we get $t_2=T$, and the sequences of pulses have trivial action on $\mathcal{H}_{\mathrm{XXX}}$ keeping it invariant. In the opposite limit, $\beta=1$, the time evolution durations become $t_0=T/2$, $t_1=T/4$, and $t_2=T/4$, and we obtain the ZX Hamiltonian.  The exact derivation of Eq.~(\ref{eqn:modulated}) is given in Appendix~\ref{sec:model_Derivation}. Additionally,   in Appendix~\ref{app:floquet_match}, we also show that for a sufficiently small $T$, this leading-order approximation accurately captures the dynamics generated by $U_T$. We emphasize that the pulse sequence adopted here is one such choices that provides the target  Hamiltonian from the original XXX Hamiltonian. We make no claim of uniqueness or optimality. Rather,  our focus is on the resulting effective Hamiltonian and the dynamics it generates. 
% Note here that though we refer to $H_{\mathrm{ZX}}$ as a graph-like Hamiltonian, unlike the cluster Hamiltonian composed of the commuting $\sigma^x_{i-1}\sigma^z_i \sigma^x_{i+1}$ stabilizers, the terms in $H_{\mathrm{ZX}}$ do not commute. Instead, the interactions $\sigma^z_{i-1} \sigma^x_{i}$ and $\sigma^x_i \sigma^z_{i+1}$ can be viewed as two-body fragments of the cluster stabilizer $\sigma^z_{i-1} \sigma^x_{i} \sigma^z_{i+1}$, with one of the neighboring $\sigma^z$ operators effectively absent. 

\begin{table}[h]
\centering
\small
\caption{Symmetries of the target Hamiltonian
$\mathcal{H}_{\mathrm{Target}}
=f_{\mathcal{H}_{\mathrm{XXX}}}(\beta)\mathcal{H}_{\mathrm{XXX}}
+f_{\mathcal{H}_{\mathrm{ZX}}}(\beta) \mathcal{H}_{\mathrm{ZX}},$
for different values of $\beta$ and impurity strength $\epsilon$.}
\label{tab:symmetries}
\begin{tabular}{p{3cm}|c|c|c|c|c|c}
\hline\hline
& \multicolumn{2}{c|}{$\beta=0$}
& \multicolumn{2}{c|}{$0<\beta<1$}
& \multicolumn{2}{c}{$\beta=1$} \\
\cline{2-7}
\textbf{Symmetry}
& $\epsilon=0$ & $\epsilon\neq0$
& $\epsilon=0$ & $\epsilon\neq0$
& $\epsilon=0$ & $\epsilon\neq0$ \\
\hline
Inversion symmetry
& \cmark & \cmark & \cmark & \cmark & \cmark & \cmark \\

$SU(2)$ symmetry
& \cmark & \xmark & \xmark & \xmark & \xmark & \xmark \\
$U(1)$ symmetry
& \cmark & \cmark & \xmark & \xmark & \xmark & \xmark \\
Chiral symmetry
& \xmark & \xmark & \xmark & \xmark & \cmark & \cmark \\

$\prod_i H_i$
& \cmark & \xmark & \cmark & \xmark & \cmark & \xmark \\

$\prod_i \sigma_i^y$
& \cmark & \xmark & \cmark & \xmark & \cmark & \xmark \\

$\sigma_i^z\sigma_{i+1}^x\sigma_{i+2}^z\cdots$
& \cmark & \xmark & \cmark & \xmark & \cmark & \cmark $^\dagger$ \\

$\sigma_i^x\sigma_{i+1}^z\sigma_{i+2}^x\cdots$
& \cmark & \xmark & \cmark & \xmark & \cmark & \cmark $^\ddagger$ \\
\hline\hline
\end{tabular}
\label{table:1}
\footnotesize
$^\dagger$ Present only when the middle site is even.\\
$^\ddagger$ Present only when the middle site is odd.

\end{table}

\section{Symmetry manipulation}
\label{sec:symmetry}

A key feature of pulse-based Hamiltonian engineering is that coherent driving can modify the symmetry structure of the target Hamiltonian $\mathcal{H}_{\text{Target}}$. Depending on the pulse parameters $\beta$ and the impurity strength $\varepsilon$, the effective model can preserve, break, or generate emergent symmetries that are not present in the original system. This allows for controlled tuning of the symmetry sectors, which is important for studying spectral statistics, quantum chaos, and eigenstate structure. In Table~\ref{table:1}, we summarize the relevant symmetries of $\mathcal{H}_{\text{eff}}$ for different parameter regions.

\begin{figure*}
    \centering
    \includegraphics[width=0.35\linewidth]{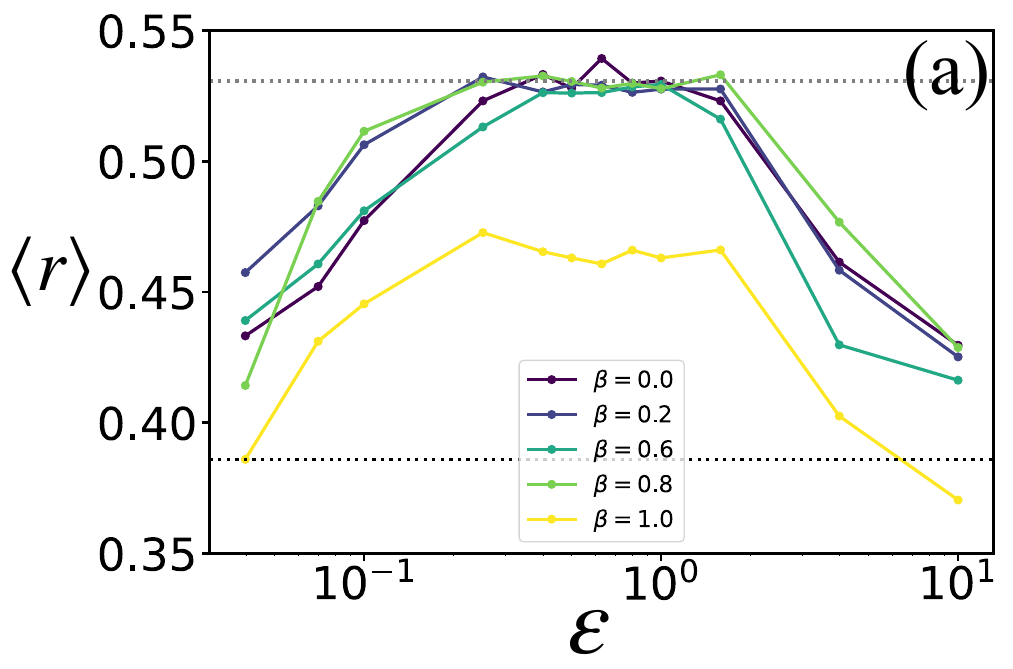}
    \includegraphics[width=\linewidth]{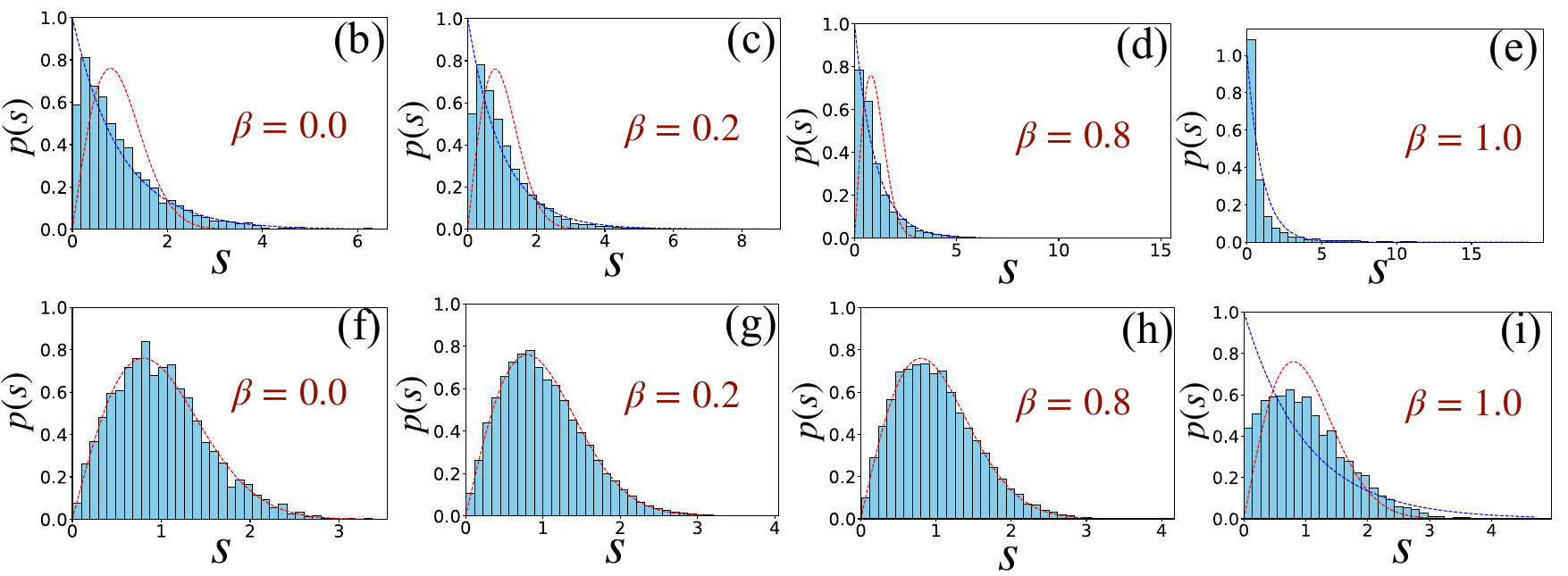}
\caption{
\textbf{Level statistics under pulse engineering.}
(a) Mean level-spacing ratio $\langle r \rangle$ as a function of impurity strength $\varepsilon$ for different values of the parameter  $\beta$. The dashed lines indicate the Poisson limit ($\langle  r \rangle_{\mathrm{Poiss}} \approx 0.38629$) and the GOE limit ($\langle r \rangle_{\mathrm{GOE}} \approx 0.53590$). In the XXX limit ($\beta=0$), the system shows Poisson-like statistics at weak impurity and crosses over to GOE behavior at stronger $\varepsilon$. For finite $\beta$ (except $\beta=1$), a similar crossover is observed, indicating that pulse engineering preserves the non-chaotic to chaotic transition across parameters. However, $\beta=1$ limits shows non-chaotic behavior for all impurity strength considered. 
Here, (b)–(e) show level spacing distribution for $\varepsilon=0.03981$ (weak impurity), while  (f)–(i) correspond to $\varepsilon=0.63095$ (strong impurity), illustrating the transition from Poisson-like to GOE-like distributions (except for $\beta=1$). Results are presented for a system of size $L=15$, restricted to the inversion-symmetric sector for all values of $\beta$. For $\beta=0$, we further restrict the calculations to the  sector with excitation-number $N_{\mathrm{ex}}=7$. For $\beta=1$, the calculations are performed in the $+1$ eigensector of the symmetry operator $\prod_i\sigma_i^z\sigma_{i+1}^x\sigma_{i+2}^z\cdots$.} 
\label{fig:level_spacing}
\end{figure*}

\begin{figure*}
    \centering
    \includegraphics[width=\linewidth]{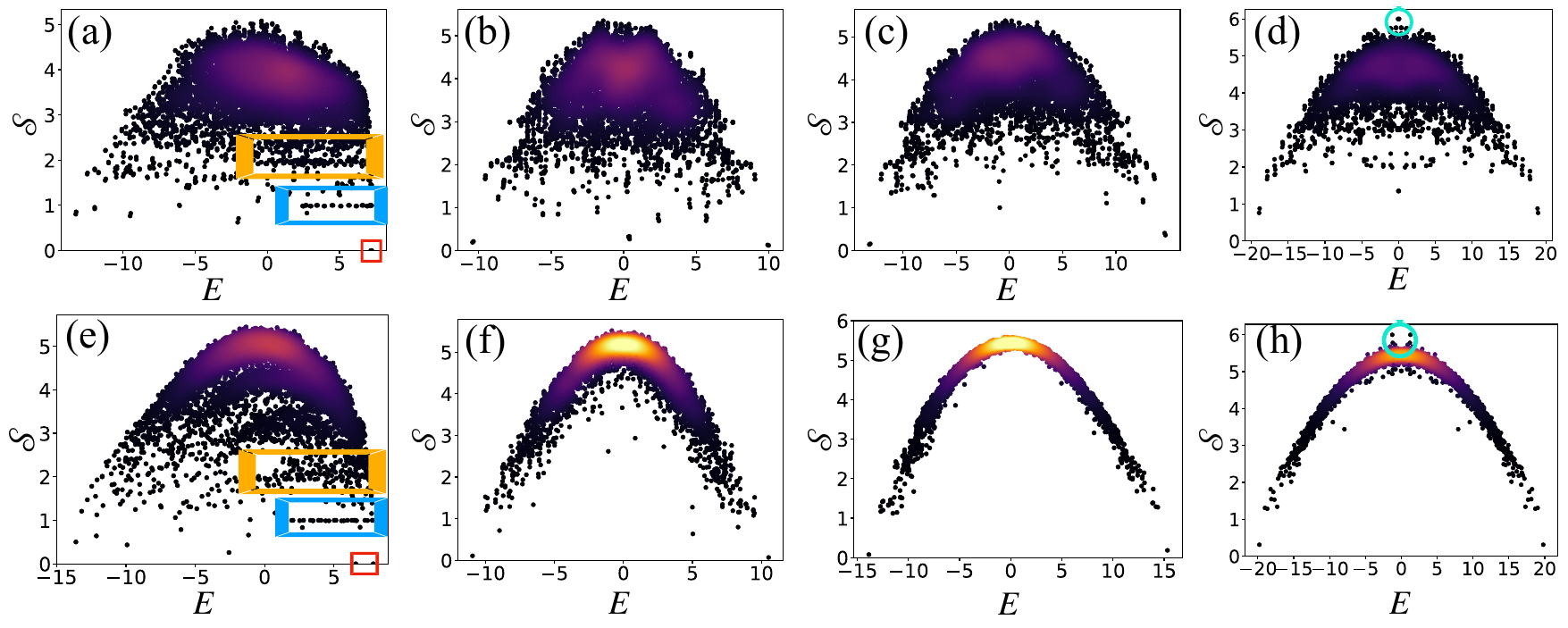}
    \caption{{\bf Energy-resolved distribution of the bipartite entanglement entropy $\mathcal{S}$ for eigenstates of the pulse-engineered spin chain}.  (a)–(d) Correspond to $\beta=0$, $0.3$, $0.7$, and $1.0$, respectively, with $\varepsilon=0.03981$. (e)–(h) Show the corresponding results for the same values of $\beta$ at $\varepsilon=0.63095$. The figures reveal distinct entanglement structures that emerge across the pulse-engineered interpolation. The color scale represents the density of states. Marked boxes of different colors highlight selected energy windows from which  atypical eigenstates are chosen for further analysis. Among them the middle one of (a) and (e) (blue box)  captures $L-1$ exact atypical states that we explore in detail. Similarly, the circle shown in (d) and (h) encloses the stabilizer scar states in the bulk of the ZX Hamiltonian. Here, we consider $L=13$.}
    \label{fig:ent_ee}
\end{figure*}
\section{Main results}

In this section, we elaborate on the main findings about  the quantum  properties of the pulse-engineered Hamiltonians. We first characterize the evolution of spectral chaos across the interpolation parameter $\beta$, and then examine the  behavior of complexity measures, and the emergence of atypical eigenstates in different parameter regimes. In the later part of our analysis, we study the time dynamics and investigate the decoupling between quantum resources generation and chaos in the pulse-engineered models.

\subsection{Chaotic properties of the pulse-engineered Hamiltonians}
\label{sec:chaos}
 To probe spectral chaos, we compute the mean level-spacing ratio
\begin{equation}
 r_n=
\frac{\min(\delta_n,\delta_{n+1})}
{\max(\delta_n,\delta_{n+1})},
\qquad
\delta_n=E_{n+1}-E_n ,
\end{equation}
and evaluate its spectral average $\langle r \rangle$. Integrable systems follow Poisson statistics with
$
\langle r \rangle_{\mathrm{Pois}}\approx0.38629,
$
while chaotic systems described by the Gaussian orthogonal ensemble satisfy
$
\langle  r \rangle_{\mathrm{GOE}}\approx0.53590.$ In our analysis, with $J=0.6$, we consider two representative impurity regimes throughout the subsequent plots: a low-impurity regime with $\varepsilon=0.03981$ and a high-impurity regime with $\varepsilon=0.63095$.
As shown in Fig.~\ref{fig:level_spacing}(a), all intermediate values of $\beta$ (except $\beta=1$) show clear level repulsion and remain close to the prediction of GOE for a  range of parameters. %{\color{cyan}Moreover, within the explored range of $\varepsilon$, the average level-spacing ratio $\langle r\rangle$ remains close to the GOE value over a broad parameter region. As a representative example, around $\varepsilon\approx0.63095$, $\langle r\rangle$ attains a high  value ??, indicating particularly strong spectral chaos at this parameter point.}  
This is supported by the level spacing distribution plot shown for some $\beta$ values in Fig.~\ref{fig:level_spacing}(b)-(d) and (f)-(h). The level spacing distribution  for $\beta=1$ for low and high impurity regimes are  shown in Figs.~\ref{fig:level_spacing}(e) and  (i), respectively.  These observations show that the pulse protocol preserves a significant amount of chaotic structure in the many-body spectrum similar to that we have in the original XXX model.

\subsection{Emergence of  atypical states}
\label{sec:static}
While the level statistics provide information about the global spectral properties of the system, they do not reveal the structure of individual eigenstates. We therefore now turn to an eigenstate-resolved analysis of the pulse-engineered Hamiltonians. As we show below, the interpolation parameter $\beta$ gives rise to three distinct regimes, each characterized by different  quantum complexity properties, and  atypical eigenstates. To characterize these regimes, we employ entanglement and nonstabilizerness as the primary diagnostics, which we briefly define in the beginning of next section.

\subsubsection{\textbf{Entanglement and magic plateaus in the $\beta=0$ (XXX) limit}}

We begin our eigenstate-resolved analysis in the limit $\beta=0$. Although, spectral statistics shown in Fig.~\ref{fig:level_spacing} indicates emergence of chaotic behavior for moderately high impurity regime, a closer examination of the eigenstates reveals the presence of atypical states that are clearly separated from the other bulk states. These states appear as distinct plateau-like structures in both entanglement and nonstabilizerness. To characterize them, we employ the measures bipartite entanglement entropy and the stabilizer Rényi entropy, which we briefly define below. 

For a pure quantum state, bipartite entanglement is quantified by the von Neumann entropy of the reduced density matrix,
\begin{equation}
\mathcal{S}
=
-\mathrm{Tr}
\left(
\rho_A\log_2\rho_A
\right),
\end{equation}
where $\rho_A=\mathrm{Tr}_{\bar A}(|\psi\rangle\langle\psi|)$ is the reduced density matrix of subsystem $A$. Unless stated otherwise, all entanglement entropy calculations presented in this work are performed for an equal bipartition of the system, i.e., $1,\dots, (L-1)/2:(L+1)/2, \dots, L$.

To quantify nonstabilizerness, we compute the stabilizer R\'enyi entropy (SRE)~\cite{Leone_2022}. The SRE for a $L$-qubit pure state $|\psi\rangle$ is defined as
\begin{equation}
{\mathcal M}(|\psi\rangle)
=-\log_2\left(\frac{1}{2^L}\sum_P
\langle \psi|P|\psi\rangle^4\right),
\end{equation}
where the sum runs over all Pauli strings $P$ of length $L$. This quantity vanishes for stabilizer states and increases with the amount of Non-Clifford structure in the state. \\

\begin{figure*}
    \centering
    \includegraphics[width=\linewidth]{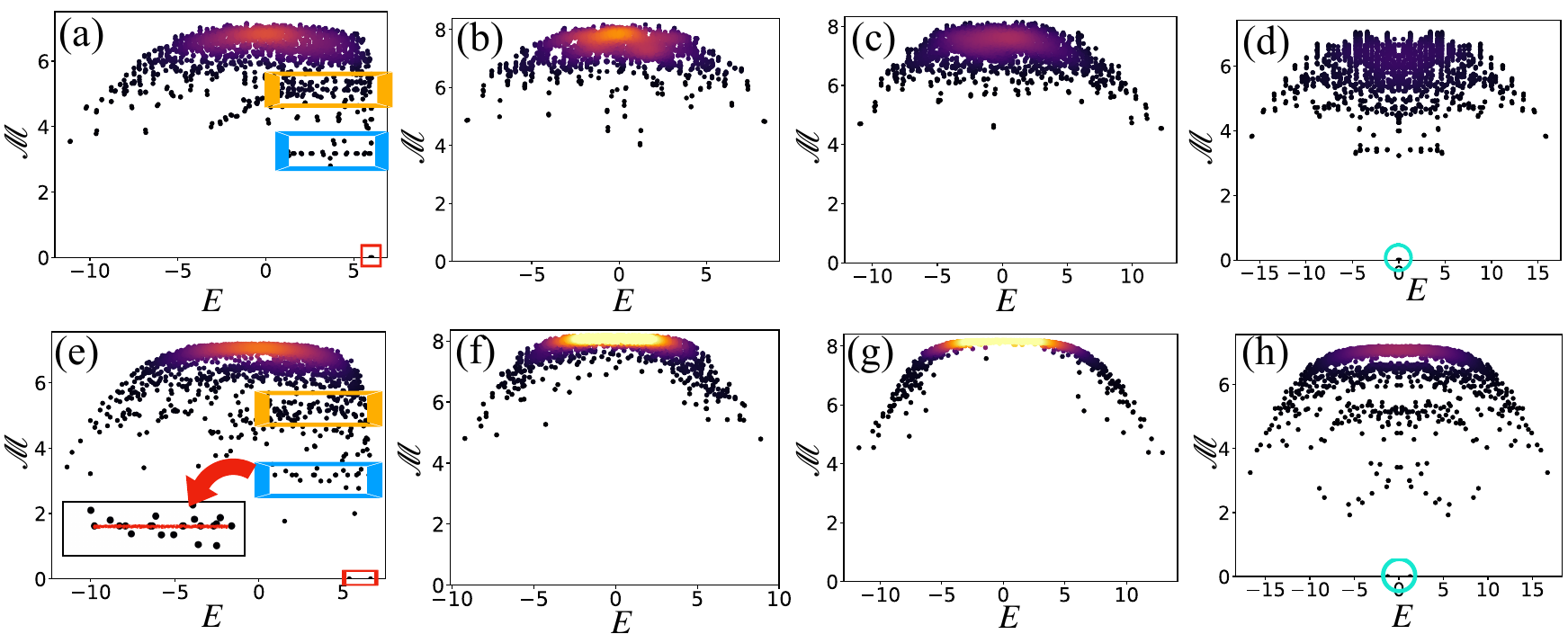}
    \caption{{\bf Energy-resolved distribution of the second-order stabilizer Rényi entropy $\mathcal{M}$ for eigenstates of the pulse-engineered spin chain}.  (a)–(d) Correspond to $\beta=0$, $0.3$, $0.7$, and $1.0$, respectively, with $\varepsilon=0.03981$. (e)–(h) Show the corresponding results for the same values of $\beta$ at $\varepsilon=0.63095$. The figures reveal distinct nonstabilizer structures that emerge across the pulse-engineered interpolation. The color scale represents the density of states. In this case, the marked boxes highlight the similar set of states  that we consider in Fig.~\ref{fig:ent_ee} for further analysis. Here, we consider $L=11$. }
    \label{fig:sre_ee}
\end{figure*}

{\bf (a) Exact atypical states with entanglement and magic plateau}: We first present the behavior of the half-chain entanglement obtained for all $2^L$ states (shown for $L=13$) in Fig.~\ref{fig:ent_ee}(a) ($\varepsilon=0.03981$), and Fig.~\ref{fig:ent_ee}(e) ($\varepsilon=0.63095$). Similarly, the behavior of nonstabilizerness (for $L=11$) is shown in Fig.~\ref{fig:sre_ee}(a) ($\varepsilon=0.03981$), and Fig.~\ref{fig:sre_ee}(e) ($\varepsilon=0.63095$). Although from the figure, one can see that the majority of eigenstates in the XXX limit ($\beta=0$) exhibit a largely featureless distribution, we observe distinct horizontal plateau-like structures in both bipartite entanglement entropy and nonstabilizerness within the highly excited part of the spectrum.  In  particular, the half-chain entanglement entropy takes the exact value  $\mathcal{S}=1$, independent of system size, while the corresponding SRE ($\mathcal{M}$) varies non-trivially with $L$.

\begin{figure*}
    \centering
    \includegraphics[width=0.7\linewidth]{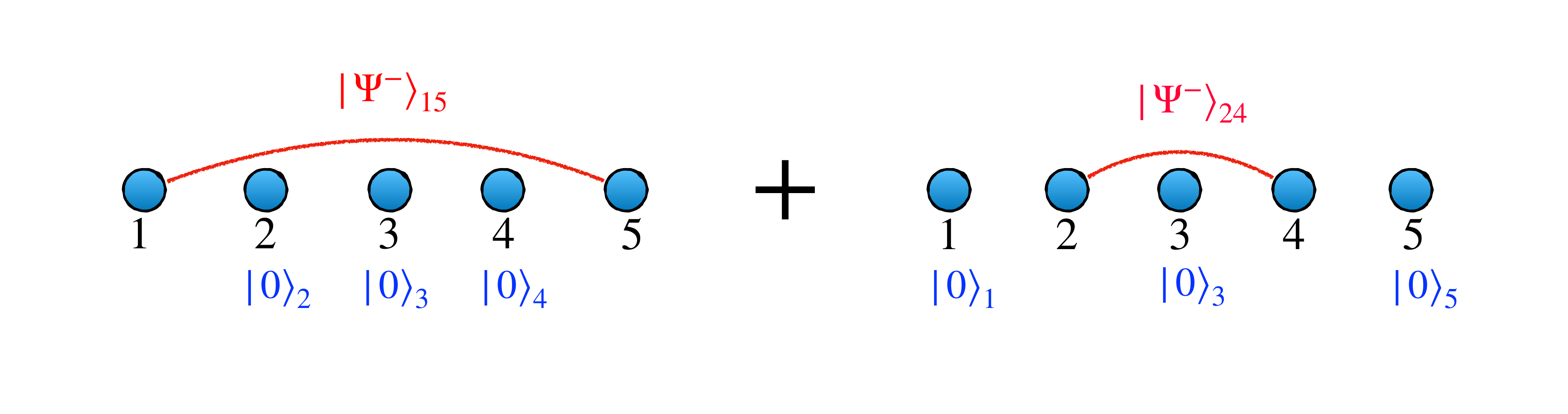}
    \caption{
{\bf Schematic representation of one of the special plateau eigenstates for $L=5$}. The state consists of a coherent superposition of different long-range Bell-pair coverings connecting distant lattice sites, together with fixed product-state configurations on the remaining spins. Such structures resemble RVB-like singlet superpositions. }
    \label{fig:sch_platue}
\end{figure*}
%To understand their origin, we analyze the corresponding eigenstates for small system sizes and extend the results up to $L=13$. 
A key observation is that the number of such special states is found to be $L-1.$   Remarkably, we are able to derive their exact analytical forms for any arbitrary system size. In the following, we present several representative examples and subsequently use an effective Hamiltonian description to derive their generic structures.

{\bf i) $L=3$:} In the smallest system ($L=3$), we have $L-1=2$ such states, and they take the form
\begin{align}
|\psi_1\rangle &= |1\rangle_2 |\Psi^-\rangle_{13}, \nonumber\\
|\psi_2\rangle &= |0\rangle_2 |\Psi^-\rangle_{13},
\end{align}
where the subscripts denote the lattice site indices, with $|\Psi^-\rangle_{kl}=\frac{|01\rangle_{kl}-|10\rangle_{kl}}{\sqrt{2}}$,
and  $\ket{0}$ and $\ket{1}$ denote the eigenstates of the operator $\sigma^z$ (the computational basis states). These states possess vanishing stabilizer R\'enyi entropy, $\mathcal{M}=0$.
Thus, these states are maximally entangled (in the bipartition $\frac{L-1}{2}:\mathrm{rest}$)  exact stabilizer states. \\

{\bf ii) $L=5$:} For the next larger system with odd sites, the same entanglement and SRE plateau persists. However,   starting from this system size, although the bipartite entanglement remains exactly $\mathcal{S}=1$, the states no longer remain stabilizer. For example, at  $L=5$, $\mathcal{M} = 1.25154$. In this case, there are four such eigenstates given by
\begin{widetext}
\begin{align}
|\psi_{1}\rangle &=
\ket{1}_3 \left(
a \ket{\Psi^-}_{15}\ket{11}_{24}
- b \ket{\Psi^-}_{24}\ket{11}_{15}
\right), \label{st1}\\
|\psi_{2}\rangle &=
\ket{0}_3 \left(
a \ket{\Psi^-}_{15}\ket{00}_{24}
- b \ket{\Psi^-}_{24}\ket{00}_{15}
\right),  \label{st2}\\
|\psi_{3}\rangle &=
\ket{1}_3 \left(
b \ket{\Psi^-}_{15}\ket{11}_{24}
+ a \ket{\Psi^-}_{24}\ket{11}_{15}
\right),  \label{st3}\\
|\psi_{4}\rangle &=
\ket{0}_3 \left(
b \ket{\Psi^-}_{15}\ket{00}_{24}
+ a \ket{\Psi^-}_{24}\ket{00}_{15}
\right), \label{st4}
\end{align}
where $a=\sqrt{\frac{5-\sqrt5}{10}}$, $b=\sqrt{\frac{5+\sqrt5}{10}}.$
\end{widetext}
%$a=0.52573$ and $b=0.85066$.

The structure of these eigenstates is reminiscent of valence-bond solid-type wavefunctions, where the many-body state is formed from coherent superpositions of different long-range singlet coverings.   A schematic representation of the states is shown in  Fig.~\ref{fig:sch_platue}.   Although their half-chain bipartite entanglement entropy remains fixed at $\mathcal{S}=1$, the underlying wavefunctions contain superposition of non-trivial long-range Bell-pair correlations extending across distant parts of the chain together with increasing nonstabilizer character as the system size grows.

{\bf (b) Effective Hamiltonian description in the exact atypical state subspace:} 
The XXX Hamiltonian conserves the total magnetization for all values of $\varepsilon$. The apparent structure of the $L-1$ special states suggests that the corresponding invariant subspace should lie entirely within either the single-excitation or the $(L-1)$-excitation sector. Consequently, by constructing the effective Hamiltonian restricted to these sectors, we derive the exact analytical forms of these states. Towards that aim, we first consider the case $\varepsilon=0$. As an illustrative example, we take $L=5$ and set $J=1$ here. Let us now  focus on the single-excitation sector, whose basis states are $|10000\rangle$, $|01000\rangle$, $|00100\rangle$, $|00010\rangle$, and $|00001\rangle$. We then combine the first two and the last two basis states into the following Bell-pair basis:

\begin{eqnarray}
    |\phi_1\rangle &=& \frac{1}{\sqrt{2}} \left( |00001\rangle - |10000\rangle \right) = |\Psi^-\rangle_{15} |000\rangle_{234}, \nonumber \\
    |\phi_2\rangle &=& \frac{1}{\sqrt{2}} \left( |00010\rangle - |01000\rangle \right) = |\Psi^-\rangle_{24} |000\rangle_{135}. \nonumber \\
\end{eqnarray}

\noindent Due to an odd number of sites, we have to leave the middle site excited state out of this calculation.  We observe that each of the $|\phi_k\rangle$ has one Bell pairing connecting the mirror symmetric sites between two halves of the system. All other sites are $|0\rangle$. Since these states are made of single excitation states, and the Hamiltonian is magnetization conserving, it turns out that these states $|\phi_1\rangle$ and $|\phi_2\rangle$ form an invariant subspace of the XXX Hamiltonian. The action of $\mathcal{H}_{\mathrm{XXX}}$ on these two states are
\begin{eqnarray}
\mathcal{H}_{\mathrm{XXX}}|\phi_1\rangle &=& 2|\phi_1\rangle + 2 |\phi_2\rangle, \nonumber \\ \mathcal{H}_{\mathrm{XXX}}|\phi_2\rangle &=& 2|\phi_1\rangle.
\end{eqnarray}
Therefore, the effective Hamiltonian in this subspace becomes
\begin{eqnarray}
    \mathcal{H}_{\mathrm{XXX}}^{\text{eff}}=
    \begin{pmatrix}
    2 & 2\\
    2 & 0
\end{pmatrix},
\end{eqnarray}
with eigenvalues $E_1= 1-\sqrt{5}, \ E_2 = 1+ \sqrt{5}$. The eigenstates of this Hamiltonian are exactly the $\frac{(L-1)}{2}=2$ atypical eigenstates with  entanglement entropy $\mathcal{S}=1$, as defined in Eqs.~(\ref{st2}) and (\ref{st4}). %The basis states $|\phi_1\rangle$ and $|\phi_2\rangle$ have exactly one Bell pair. \textcolor{red}{This information is somehow getting translated into the eigenstates also, where, although there are two Bell pairs, the coefficients adjust them to produce entanglement entropy exactly $1$}. \\

\noindent A pattern of the $\mathcal{H}_{\mathrm{XXX}}^{\text{eff}}$ emerges when we look at higher $L$ also. For instance, for $L=7$, we can construct $3$ such basis vectors $|\phi_1\rangle, |\phi_2\rangle, |\phi_3\rangle$. The action of the Hamiltonian on these basis state is given by
\begin{eqnarray}
   \mathcal{H}_{\mathrm{XXX}}|\phi_1\rangle &=& 4 |\phi_1\rangle + 2|\phi_2\rangle, \nonumber \\ \mathcal{H}_{\mathrm{XXX}}|\phi_2\rangle &=& 2|\phi_1\rangle + 2|\phi_2\rangle + 2|\phi_3\rangle,\nonumber \\ \mathcal{H}_{\mathrm{XXX}}|\phi_3\rangle &=& 2 |\phi_2\rangle + 2|\phi_3\rangle. 
\end{eqnarray} Therefore, in that basis $\mathcal{H}_{\mathrm{XXX}}^{\text{eff}}$ becomes
\begin{eqnarray}
    \mathcal{H}_{\mathrm{XXX}}^{\text{eff}}=
    \begin{pmatrix}
    4 & 2 & 0\\
    2 & 2 & 2\\
    0 & 2 & 2\\
\end{pmatrix}.
\end{eqnarray}

\noindent Defining $N= (L-1)/2$, the general action can be written as follows: 
\begin{align}
    &\mathcal{H}_{\mathrm{XXX}}|\phi_1\rangle = (2N-2) |\phi_1\rangle + 2|\phi_2\rangle,& \nonumber \\
    &\mathcal{H}_{\mathrm{XXX}}|\phi_n\rangle = 2|\phi_{n-1}\rangle + (2N-4)|\phi_n\rangle + 2|\phi_{n+1}\rangle,& \nonumber \\
    &\mathcal{H}_{\mathrm{XXX}}|\phi_N\rangle = 2 |\phi_{N-1}\rangle + (2N-4)|\phi_N\rangle.&
\end{align}

\noindent Here, $n=2,3, ...,N-1$. Therefore, the general form of $\mathcal{H}_{\text{XXX}}^{\text{eff}}$ can be written as
\begin{eqnarray}
    \mathcal{H}_{\mathrm{XXX}}^{\text{eff}}=
    \begin{pmatrix}
    2N-2 & 2 & 0 & 0 & \cdots & 0 \\
    2 & 2N-4 & 2 & 0 & \cdots & 0\\
    0 & 2 & 2N-4 & 2 & \cdots & 0\\
    \vdots & \vdots & \vdots & \vdots & \cdots & \vdots \\
    0 & 0 & 0 & 0 &\cdots  & 2N-4\nonumber\\
\end{pmatrix}.
\end{eqnarray}
\noindent  We can further write  $\mathcal{H}_{\mathrm{XXX}}^{\text{eff}}$ as $(2N-4) \mathbb{I} + K$ where $K$ is given by 
\begin{eqnarray}
    K=
    \begin{pmatrix}
    2 & 2 & 0 & 0 & \cdots & 0 \\
    2 & 0 & 2 & 0 & \cdots & 0\\
    0 & 2 & 0 & 2 & \cdots & 0\\
    \vdots & \vdots & \vdots & \vdots & \cdots & \vdots \\
    0 & 0 & 0 & 0 &\cdots  & 0
\end{pmatrix}.
\end{eqnarray}

\noindent  Now consider that the eigenvalue equation is given by $\mathcal{H}_{\mathrm{XXX}}^{\text{eff}} \ \psi = E \psi$. Then we must have $K \psi = \lambda \psi$ and $E= (2N-4) + \lambda$. Let us start with the general form of the eigenstates,  $\psi = [\psi_1 \ \psi_2 \ \cdots \psi_N]^T$. For $n=2, 3,..., N-1$, from the equation $K\psi=E\psi$, we get
\begin{align}
    &2 \psi_{n-1} + 2\psi_{n+1} = \lambda \psi_n& \nonumber 
    \Rightarrow \ &\psi_{n-1} + \psi_{n+1} = \frac{\lambda}{2} \psi_n.&
\end{align}

\noindent To solve it, we consider the trial wavefunction $\psi_n = e^{-ikn}$. This leads to
\begin{align}
    &e^{ik} + e^{-ik} = \frac{\lambda}{2}& \nonumber 
    \Rightarrow \ &\lambda = 4\cos k.
\end{align}

\noindent Since the ansatz $\psi_n=e^{-ikn}$ leads to the same eigenvalue, the most general solution can be written as
\begin{align}
    \psi_n &= A^{\prime} e^{ikn} + B^{\prime} e^{-ikn}, \nonumber \\
  &= A \sin kn + B \cos kn. 
    \label{psi_n}
\end{align}

\noindent From the last row of $K$, we get 
\begin{align}
    2 \psi_{N-1} = \lambda \psi_N = 4\cos k\  \psi_N
    \label{psi_N}.
\end{align}

\noindent Putting $\psi_n$ from Eq.~\eqref{psi_n} into Eq.~\eqref{psi_N} and doing some simplifications, we get
\begin{align}
   A \sin[k(N+1)] + B \cos[k(N+1)]=0, 
   \label{psi_N+1}
\end{align}

\noindent which is nothing but the condition $\psi_{N+1}=0$. From Eq.~\eqref{psi_N+1}, we further get $B = -A \tan[k(N+1)]$. Therefore, 
\begin{align}
    \psi_n &= A \sin kn \ - A \tan[k(N+1)] \cos kn & \nonumber \\ 
     & =\frac{-A}{\cos[k(N+1)]} \Big[\sin[k(N+1)] \cos kn -\nonumber \\
     &\cos [k(N+1)]\sin kn\Big],& \nonumber \\
    & = \frac{A}{\cos[k(N+1)]} \ \sin[k(N+1-n)], \end{align}
\noindent where we have absorbed the minus sign in the constant $A$. From the first row of $K$, we get
\begin{align}
    &2\psi_1 + 2\psi_2 = \lambda\psi_1, 
    \ \Rightarrow \psi_2 = (2\cos k -1) \psi_1.
\end{align}

\noindent Inserting $\psi_1 = \frac{A}{\cos[k(N+1)]} \sin Nk$ and $\psi_2 = 
\frac{A}{\cos[k(N+1)]} \sin[(N-1)k]$ and doing some algebra, we get
\begin{align}
    &\tan Nk = \cot \frac{k}{2}  \Rightarrow  Nk = \frac{\pi}{2} - \frac{k}{2} + m\pi, \nonumber \\
    \Rightarrow \ & k_m = \frac{(2m+1)\pi}{2N+1}, \ \ \ m= 0, 1, ..., N-1.
\end{align}

\noindent Hence, the energy eigenvalues takes the following exact analytical form are given by 
\begin{align}
    E_m &= (2N-4) + 4\cos \left( \frac{(2m+1)\pi}{2N+1}\right),& \nonumber \\
     &= (L-5) + 4\cos\left( \frac{(2m+1)\pi}{L} \right).
\end{align}
\begin{figure}
    \centering
    \includegraphics[width=0.75\linewidth]{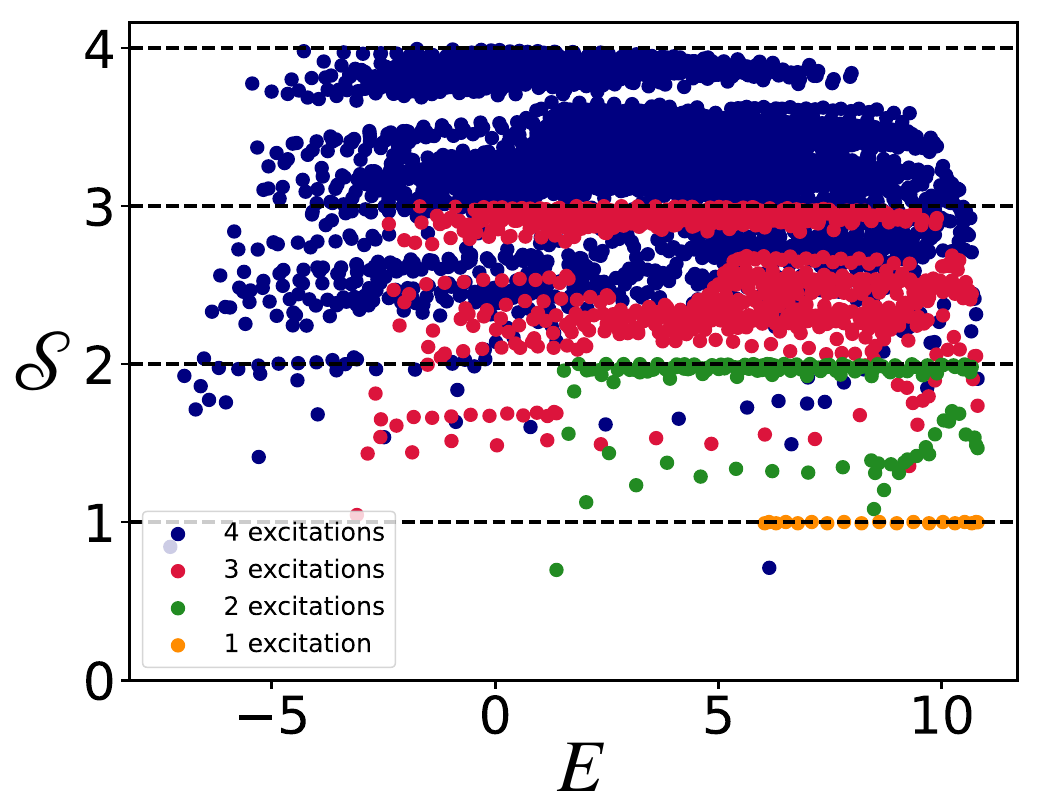}
\caption{{\bf Half-chain entanglement entropy of the approximate atypical eigenstates of the chaotic XXX chain.}  The orange points correspond to the exact $L-1$ plateau states in the single-excitation sector with $\mathcal{S}=1$. The green, red, and blue points denote the approximate atypical states in the two-, three-, and four-excitation sectors, respectively. These states form approximate entanglement plateaus around entanglement entropy values $\mathcal{S}\approx2$, $3$, and $4$, illustrating a hierarchy of atypical states with nearly quantized entanglement values present together with   the exact plateau state. Here, we consider  $L=19$, $\beta=0$, and $\varepsilon=0.0$.}
    \label{fig:scaling_atypicalstates}
\end{figure}
\noindent As a result, the $n$'th element of the $m$'th eigenvector is 
\small{
\begin{align}
    \psi_n^{(m)} = \frac{A}{\cos \left[\frac{(2m+1)\pi}{L}\left(\frac{L+1}{2}\right)\right]} \sin \left[\frac{(2m+1)\pi}{L} \left( \frac{L+1}{2} -n\right) \right].
\end{align}}
\noindent Using the normalization condition $\sum_{n=1}^N |\psi_n^{(m)}|^2 = 1$, we finally get
\begin{align}
    \psi_n^{(m)} = \frac{2}{\sqrt{L}} \sin\left[ \frac{(2m+1)\pi}{L} \left( \frac{L+1}{2} -n\right)\right].
\end{align}

\noindent Hence, the above analysis helps us to obtain the explicit expressions for the   $L-1$ atypical eigenvectors and their corresponding energies, for which the entanglement entropy is exactly equal to $1$.

\begin{figure*}
    \centering
    \includegraphics[width=\linewidth]{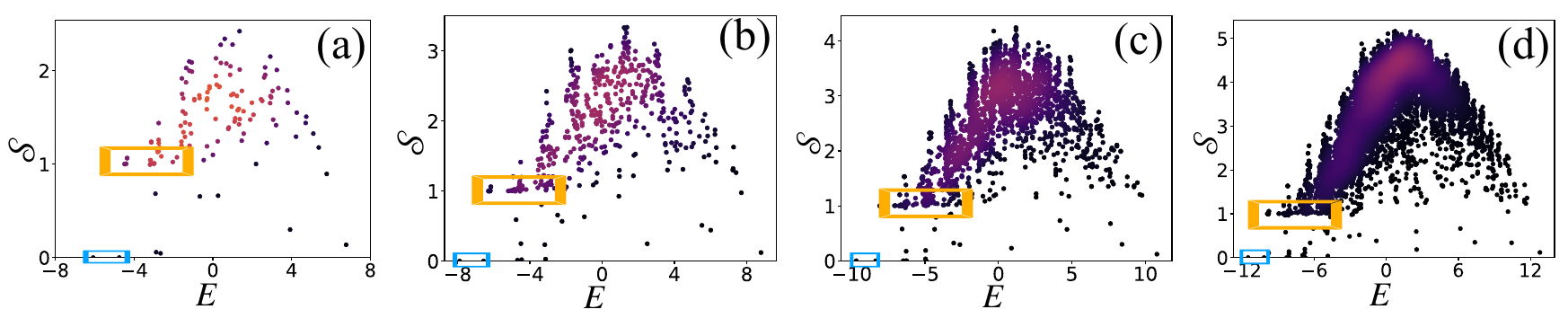}
\caption{{\bf Atypical states for intermediate $\beta$.} Energy-resolved distribution of the bipartite entanglement entropy for $\beta=0.5$ and $\varepsilon=0.63095$.  (a)–(d) correspond to $L=7$, $9$, $11$, and $13$, respectively. The color scale represents the density of states. The square boxes highlight the low-entanglement eigenstates. As the system size increases, a hierarchy of atypical eigenstates with entanglement entropy close to $\mathcal{S}\approx0$ and $1$ becomes progressively more pronounced.}
    \label{fig:ent_ee_new}
\end{figure*}
\noindent However, till now we were discussing in single excitation sector. The same calculations can be done for $L-1$ excitation sector. For $L=5$, the basis states in this subspace are given by  $|01111\rangle, |10111\rangle, |11011\rangle, |11101\rangle$, and $|11110\rangle$. As in the previous case,we get the basis states, given below. 
\begin{eqnarray}
|\xi_1\rangle = |\Psi^-\rangle_{15}|111\rangle_{234}, \ \
|\xi_2\rangle = |\Psi^-\rangle_{24}|111\rangle_{135}. 
\end{eqnarray}

\noindent The algebra is similar to the previous one, and here also,  we get the required effective Hamiltonian, yielding the states Eqs.~(\ref{st1}) and (\ref{st3}). Therefore, these two derivations together yield four eigenstates, corresponding to two distinct energy values, each having an entanglement entropy of exactly $1$.  This fact can be understood from the global spin flip or $\mathbb{Z}_2$ symmetry of the XXX Hamiltonian. We have $|\xi_1\rangle = \left( \prod_i \sigma^x_i \right) |\phi_1\rangle$ and $|\xi_2\rangle = \left( \prod_i \sigma^x_i \right) |\phi_2\rangle$ up to some global phases. 

However, these special energy values, at which these states occur are not just two-fold degenerate. The XXX Hamiltonian has in fact a bigger symmetry than just $\mathbb{Z}_2$ symmetry- it has a global $SU(2)$ symmetry. % We can define $S^z =  \frac{1}{2} \sum_i \sigma^z_i$ and $S^{\pm} = \sum_i \sigma^{\pm}$ where $\sigma^{\pm} = \frac{1}{2} (\sigma^x \pm i \sigma^y)$. Then we have $[H_{\text{XXX}}, S^z] =0,\ [H_{\text{XXX}}, S^{\pm}] =0$ and $[H_{\text{XXX}}, S^2]=0$. 
The singlet part has $S=0$, and the part containing the rest three spins has $S=3/2$.  As we know that any spin-$S$ multiplet contains $2S+1$ number of degenerate states, this is the reason  why we obtain  $4$ degenerate states here at $E_1$ and $E_2$. In general, for a system of size $L$, we find $L-1$ degenerate states at each of the $(L-1)/2$ distinct energy values. Among the degenerate states at each energy, two states have a half-chain entanglement entropy exactly equal to $1$. Thus, in total, there are $L-1$ eigenstates with half-chain entanglement entropy equal to $1$. \\

\noindent However, the middle site impurity breaks the $SU(2)$ symmetry. Therefore, each of those degenerate energies gets splitted. As in $|\phi_1\rangle$ and $|\phi_2\rangle$, we had $|0\rangle_3$ in the middle site, we now have
\begin{eqnarray}
  \varepsilon \sigma^z_3|\phi_1\rangle = +\varepsilon|\phi_1\rangle, \ \ \varepsilon \sigma^z_3|\phi_2\rangle = +\varepsilon|\phi_2\rangle . 
\end{eqnarray}

\noindent Hence, each energy eigenvalue gets a $+\varepsilon$ contribution. Similarly, for $|\xi_1\rangle, \ |\xi_2\rangle$ we have
\begin{eqnarray}
  \varepsilon \sigma^z_3|\xi_1\rangle = -\varepsilon|\xi_1\rangle, \ \ \varepsilon \sigma^z_3|\xi_2\rangle = -\varepsilon|\xi_2\rangle . 
\end{eqnarray}

\noindent As a result, the degenerate eigenstates pair with entanglement entropy $1$ at each $E_i$'s gets splitted into two separate states at energy $E_i - \varepsilon$ and $E_i + \varepsilon$. Thus, we get a total $L-1$ number of non-degenerate states with entanglement entropy $\mathcal{S}=1$ at energy values
\begin{align}
    E_m = (L-5) + 4\cos\left( \frac{(2m+1)\pi}{L} \right) \pm \varepsilon ,
\end{align}
\noindent where $m=0,1,..., \frac{L-3}{2}$. \\

\noindent We emphasize that, despite their exact analytical structure and clear deviation from thermal-state behavior, we conservatively refer to these states as exact atypical states rather than quantum many-body scars. In the literature, atypical states such as the $\eta$-pairing states of the Hubbard model~\cite{Moudgalya_2020,Daniel_2020} arise due to an underlying $SU(2)$ symmetry and are therefore generally not classified as scars. In the present case, the $L-1$ atypical states are associated with the global $U(1)$ conservation of total magnetization. Interestingly, preliminary results indicate that some of their characteristic features may survive even under weak breaking of the $U(1)$ symmetry, leaving open the question of their precise relation to quantum many-body scar physics.\\

{\bf (c) Approximate entanglement plateaus for high magnetization sectors:} 
In addition to the exact plateau states in Fig.~\ref{fig:ent_ee}(a) (orange square box), we identify a hierarchy of approximate plateaus of entanglement in different excitation sectors, with eigenstates clustered around half-chain entanglement entropy values close to $2, 3, 4 \cdots$. As the system size increases, these approximate plateaus extend to higher-excitation sectors. For a sector with excitation number $k$, the half-chain entanglement entropy is bounded by $\min(k,L-k)$. For instance, for $L=19$, Fig.~\ref{fig:scaling_atypicalstates} shows a significant number of states in the higher-excitation sectors $(2,3,4)$ with entanglement entropy values approaching the corresponding bounds, $\mathcal{S}\approx 2$, $3$, and $4$, respectively. In contrast, we find that the corresponding magic does not exhibit any clear or systematic behavior across these approximate plateaus. Interestingly, once we take $\varepsilon >0$, the entanglement entropy  bound persists only for excitation sector 1 (or $L-1$), further highlighting the special nature of the $L-1$ atypical states.

\subsubsection{\textbf{Atypical states of the pulse-engineered Hamiltonian for $0<\beta<1$}}

The behavior changes qualitatively throughout the intermediate regime $0<\beta<1$. As evident from Figs.~\ref{fig:ent_ee}(b), (c) and Figs.~\ref{fig:sre_ee}(b), (c), the isolated family of $L-1$ atypical states observed in the XXX limit is no longer present. Instead, the low-entanglement and low-stabilizerness eigenstates become more broadly distributed, indicating that the pulse-induced deformation weakens the distinct atypical structure. Interestingly, a closer examination of the intermediate regime reveals the emergence of a different class of atypical states.  As we show in Fig.~\ref{fig:ent_ee_new}, the low-energy part of the spectrum develops a hierarchy of approximate plateaus with entanglement entropy values close to $\approx 0, 1$ ~\cite{Alba_2009}. However, deriving their exact analytical form for arbitrary system sizes is extremely hard.

%To further understand the evolution of the eigenstate structure, we now examine two standard diagnostics motivated by the eigenstate thermalization hypothesis (ETH). Specifically, we study the matrix elements of the observable
%\begin{equation}
%    \hat{O}=\sum_i \sigma_i^z,
%\end{equation}  $\langle E_n|\hat{O}|E_m\rangle$ as shown in Fig.~\ref{fig:eth}(a)–(d). We further plot the  diagonal expectation values of the same observable as a function of energy in Fig.~\ref{fig:eth}(e)–(h). The figures show for $\beta=0.5$, the fluctuations in the diagonal elements with energy  decrease with increasing system size, consistent with the expected trend toward eigenstate thermalization. In contrast, the pulse-engineered ZX limit ($\beta=1$) exhibits qualitatively different behavior.

 \subsubsection{\textbf{Zero-magic maximally entangled states for \texorpdfstring{$\beta=1$}{beta=1}\\ : rainbow scars}}
The last set of example of atypical state we present is in the the ZX limit of the pulse-engineered Hamiltonian. In contrast to the $0\leq \beta<1$ case, for $\beta=1$ we observe a pair of special eigenstates in the spectrum that exhibit {\it vanishing} stabilizer R\'enyi entropy (SRE) while possessing {\it maximal} bipartite entanglement, see Figs.~\ref{fig:ent_ee}(d), (h), for bipartite entanglement  and Figs.~\ref{fig:sre_ee}(d), (h) for nonstabilizerness.  These two states occur at energies $\pm E$, with an energy separation of $4\varepsilon$. Their combination of maximal bipartite entanglement and vanishing stabilizer Rényi entropy is characteristic of the well-known \textit{rainbow  states}~\cite{Ram_rez_2014,Ram_rez_2015}. 

Below we write the exact form of the above states and later provide a general form for all $L$. 

\vspace{0.5em}
\noindent {\it $L=3$:}
\begin{align}
|\chi_1\rangle &=  \ket{\Psi^-}_{13}\ket{-}_2, \nonumber \\
|\chi_2\rangle &= \ket{\Psi^-}_{13}\ket{+}_2.
\end{align}

\noindent {\it $L=5$:}
\begin{align}
|\chi_{1}\rangle &= \ket{\Phi^+}_{15}\ket{\Psi^-}_{24}\ket{-}_3,\nonumber \\
|\chi_{2}\rangle &= \ket{\Phi^+}_{15}\ket{\Psi^-}_{24}\ket{+}_3.
\end{align}

\noindent {\it $L=7$:}
\begin{align}
|\chi_{1}\rangle &= \ket{\Psi^-}_{17}\ket{\Phi^+}_{26}\ket{\Psi^-}_{35}\ket{-}_4, \nonumber \\
|\chi_{2}\rangle &= \ket{\Psi^-}_{17}\ket{\Phi^+}_{26}\ket{\Psi^-}_{35}\ket{+}_4.
\end{align}

\noindent {\it $L=9$:}
\begin{align}
|\chi_{1}\rangle &= \ket{\Phi^+}_{19}\ket{\Psi^-}_{28}\ket{\Phi^+}_{37}\ket{\Psi^-}_{46}\ket{-}_5, \nonumber \\
|\chi_{2}\rangle &= \ket{\Phi^+}_{19}\ket{\Psi^-}_{28}\ket{\Phi^+}_{37}\ket{\Psi^-}_{46}\ket{+}_5.
\end{align}

\vspace{0.5em}

From these examples, a clear pattern emerges. For odd system size $L$ and $m=\frac{L+1}{2}$, the above two stabilizer states can be written  in the general form using the following equations
\begin{align}
|\chi_1\rangle &= \bigotimes_{k=1}^{(L-1)/2} |\mathcal{B}_k\rangle_{k,\,L+1-k} \otimes |-\rangle_m, \nonumber \\
|\chi_2\rangle &= \bigotimes_{k=1}^{(L-1)/2} |\mathcal{B}_k\rangle_{k,\,L+1-k} \otimes |+\rangle_m,
\end{align}
where the two-qubit Bell states alternate as

\begin{figure*}
\includegraphics[width=0.85\linewidth]{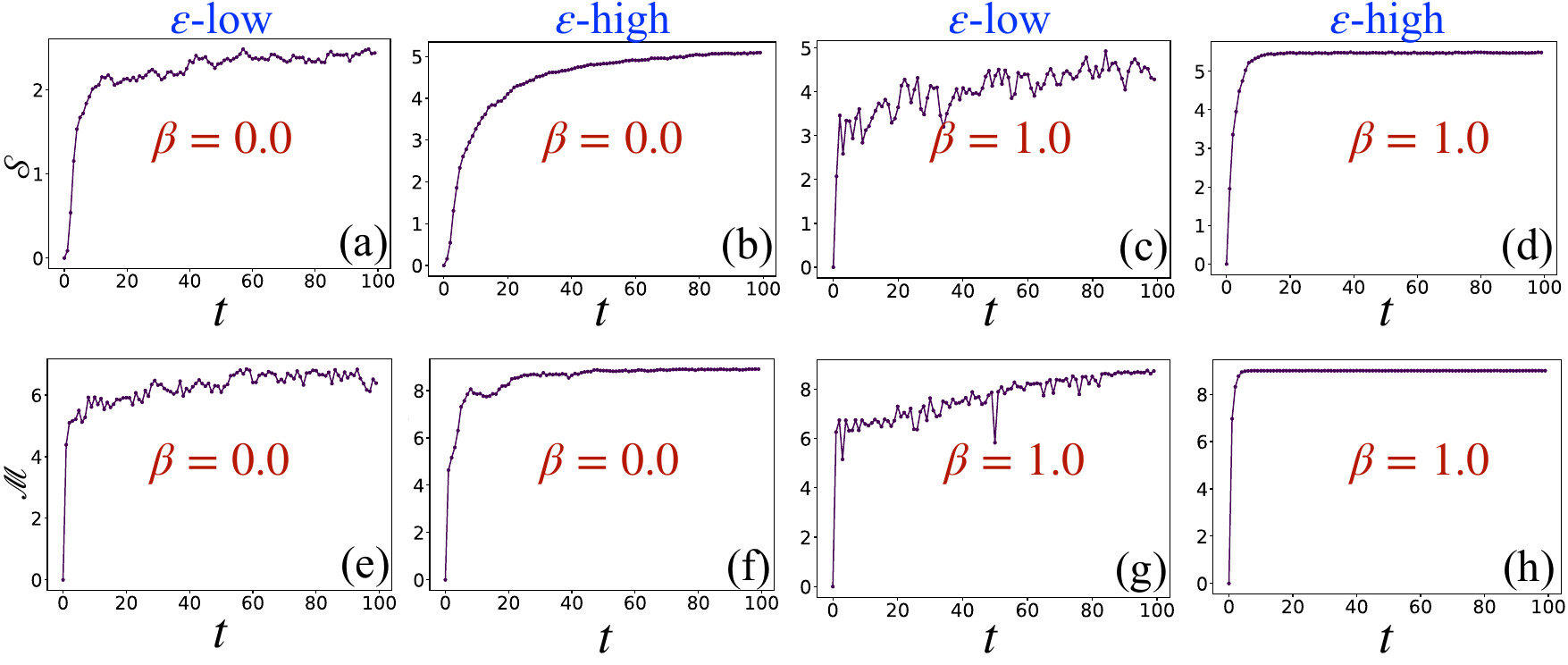}
\caption{{\bf Time evolution of the half-chain entanglement entropy $\mathcal{S}$ and stabilizer Rényi entropy $\mathcal{M}$.}
The dynamics are shown for a representative initial product stabilizer state under the  evolution generated by the pulse-engineered Hamiltonian. The upper row ((a)–(d)) shows the evolution of the half-chain entanglement entropy $\mathcal{S}$, while the lower row ((e)–(h)) shows the corresponding stabilizer Rényi entropy $\mathcal{M}$. The left halves correspond to the undeformed Hamiltonian ($\beta=0$), and the right halves to the ZX limit ($\beta=1$). Within each case, the left and right panels represent the low- ($\varepsilon=0.03981$) and high-impurity ($\varepsilon=0.63095$) regimes, respectively. For $\beta=0$, these correspond to the non-chaotic and chaotic regimes, whereas for $\beta=1$ the system remains non-chaotic for all impurity strengths. In both limits, increasing the impurity strength leads to faster growth and higher saturation values of both the entanglement entropy and the stabilizer Rényi entropy. For entanglement dynamics the  results are shown for $L=13$. Whereas, for nonstabilizerness we consider $L=11$. The chosen initial state is one representative member of the ensemble of 20 random product stabilizer states used in Fig.~\ref{fig:max_ent_sre}.}
\label{fig:ent_time}
\end{figure*}
\begin{equation}
|\mathcal{B}_k\rangle=
\begin{cases}
\begin{cases}
|\Psi^-\rangle & \text{if } k\ \text{is odd},\\
|\Phi^+\rangle & \text{if } k\ \text{is even},
\end{cases}
& \text{if } m\ \text{is even},\\[8pt]
\begin{cases}
|\Phi^+\rangle & \text{if } k\ \text{is odd},\\
|\Psi^-\rangle & \text{if } k\ \text{is even},
\end{cases}
& \text{if } m\ \text{is odd}.
\end{cases}
\end{equation}
where
\begin{align}
|\Psi^-\rangle_{kl}
&=\frac{1}{\sqrt{2}}\left(|01\rangle_{kl}-
|10\rangle_{kl}\right), \nonumber\\
|\Phi^+\rangle_{kl}&=\frac{1}{\sqrt{2}}\left(|00\rangle_{kl}
+|11\rangle_{kl}\right).
\end{align}

\begin{figure*}
\includegraphics[width=0.85\linewidth]{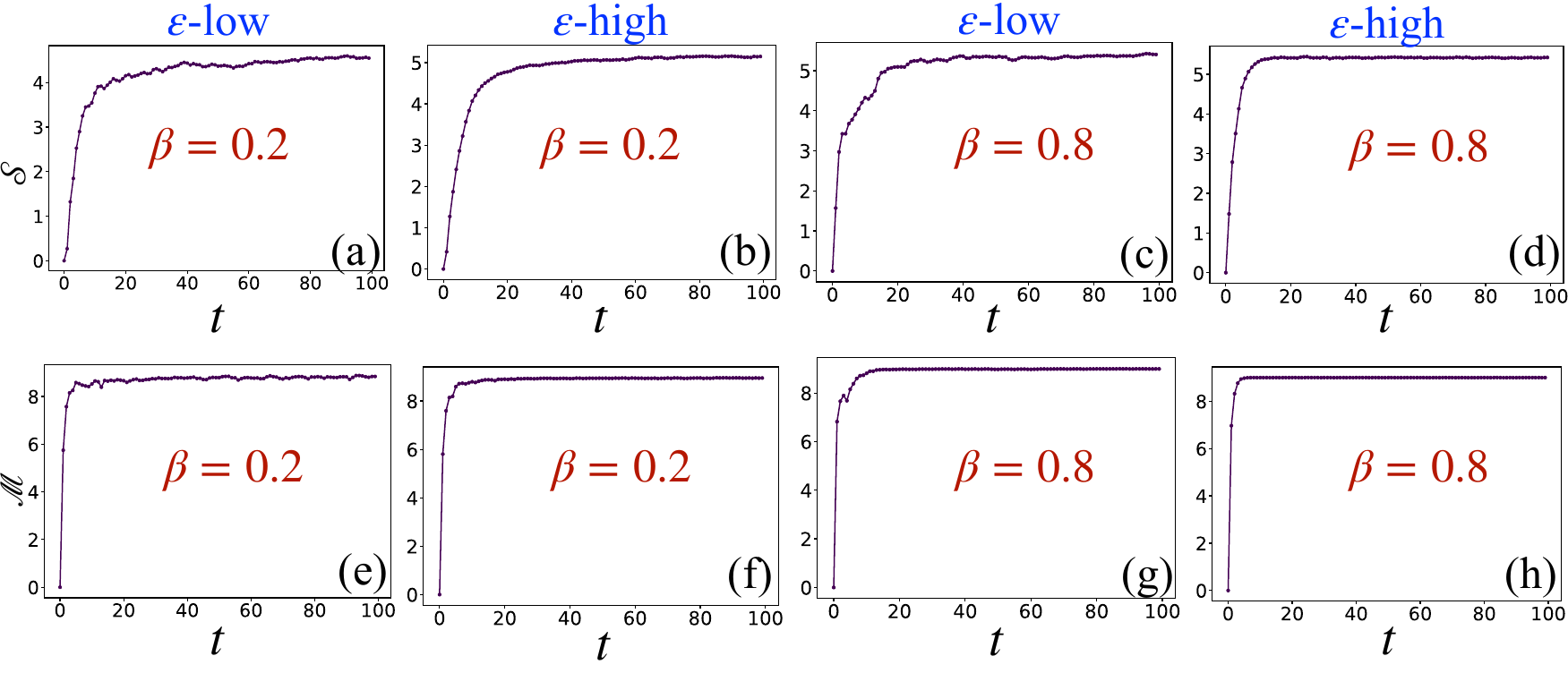}
\caption{{\bf Time evolution of the half-chain entanglement entropy $\mathcal{S}$ and stabilizer Rényi entropy $\mathcal{M}$ for intermediate pulse strengths.}
The upper row ((a)–(d)) shows the evolution of the half-chain entanglement entropy, while the lower row ((e)–(h)) shows the corresponding stabilizer Rényi entropy. The left and right halves correspond to $\beta=0.2$ and $0.8$, respectively. Within each case, the left (right) panels represent the low-($\varepsilon=0.03981$) (high-($\varepsilon=0.63095$)) impurity  regime. In contrast to the $\beta=0$ and $\beta=1$ limits shown in Fig.~\ref{fig:ent_time}, the dynamics for the low- and high-impurity regimes are nearly indistinguishable for both entanglement and magic at intermediate $\beta$. These results demonstrate that  pulse engineering substantially weakens the dependence of quantum resource generation on the underlying spectral chaos. For entanglement and magic dynamics the  results are shown for the same system sizes as considered in Fig~\ref{fig:ent_time}. The chosen initial state is one representative member of the ensemble of 20 random product stabilizer states used in Fig.~\ref{fig:max_ent_sre} .}
\label{fig:sre_time}
\end{figure*}

These states therefore correspond to a product of alternating long-range Bell pairs formed between mirror-symmetric lattice sites, together with a single central qubit in the $|\pm\rangle$ basis. Consequently, they exhibit maximal bipartite entanglement across the half-chain cut while remaining exact stabilizer states with vanishing SRE. Moreover, the half-chain entanglement entropy in this case grows linearly with the system size $L$, indicating volume-law entanglement. The same structure persists for all odd system sizes, explaining the robustness of these special eigenstates within the spectrum. Owing to their exact Bell-pair structure and long-range entanglement, such states may also be useful for quantum communication and entanglement distribution protocols.  As these states lie near the middle of the spectrum and exhibit a structure that is significantly different from the other bulk eigenstates, we identify them as rainbow scar states~\cite{Langlett_2022}. Moreover, although they show volume-law entanglement for the half-chain bipartition, an appropriate choice of bipartition reveals an area-law entanglement structure~\cite{Singha_Roy_2020}, further supporting the low entanglement structure of scar states.

\subsection{Pulse-Induced Decoupling of Quantum Resources and Chaos}
\label{sec:dynamics}
%Having established the existence and robustness of the atypical eigenstates across different pulse regimes, we now investigate their dynamical consequences. 
A central question in quantum many-body dynamics concerns the interplay between spectral chaos and the generation of quantum resources such as entanglement and nonstabilizerness. In generic interacting systems, chaotic spectra are often associated with rapid entanglement growth~\cite{Kim_2013,Nahum2017}, and increasing complexity of the evolving quantum state. Recent studies have further suggested that chaotic dynamics can also enhance the generation of nonstabilizerness~\cite{Kanato_2022,Dowling_2025,Oliviero_2025}, although the precise relation between spectral chaos and magic remains less understood than that for entanglement. This naturally raises the question of whether pulse engineering can reshape these connections and decouple the generation of quantum resources from conventional signatures of chaos. In particular, we ask whether the relation between spectral chaos and measures of state complexity remains similar to that in the original Hamiltonian. 

To this end, we consider  time evolution under the target Hamiltonian $\mathcal{H}_{\mathrm{Target}}$, starting from random stabilizer-product states $|\psi(0)\rangle_{\mathrm{Stab}}^{\mathrm{Product}}$,
\begin{equation}
|\psi(t)\rangle=
e^{-i\mathcal{H}_{\mathrm{Target}}t}
|\psi(0)\rangle_{\mathrm{Stab}}^{\mathrm{Product}}.
\label{eqn:evolution}
\end{equation}
 Since stabilizer states do not have magic, they provide a convenient reference for quantifying their dynamical generation. %Unless stated otherwise, {\color{blue}all results are averaged over $20$ such initial states}.

\begin{figure*}

        \includegraphics[width=0.6\linewidth]{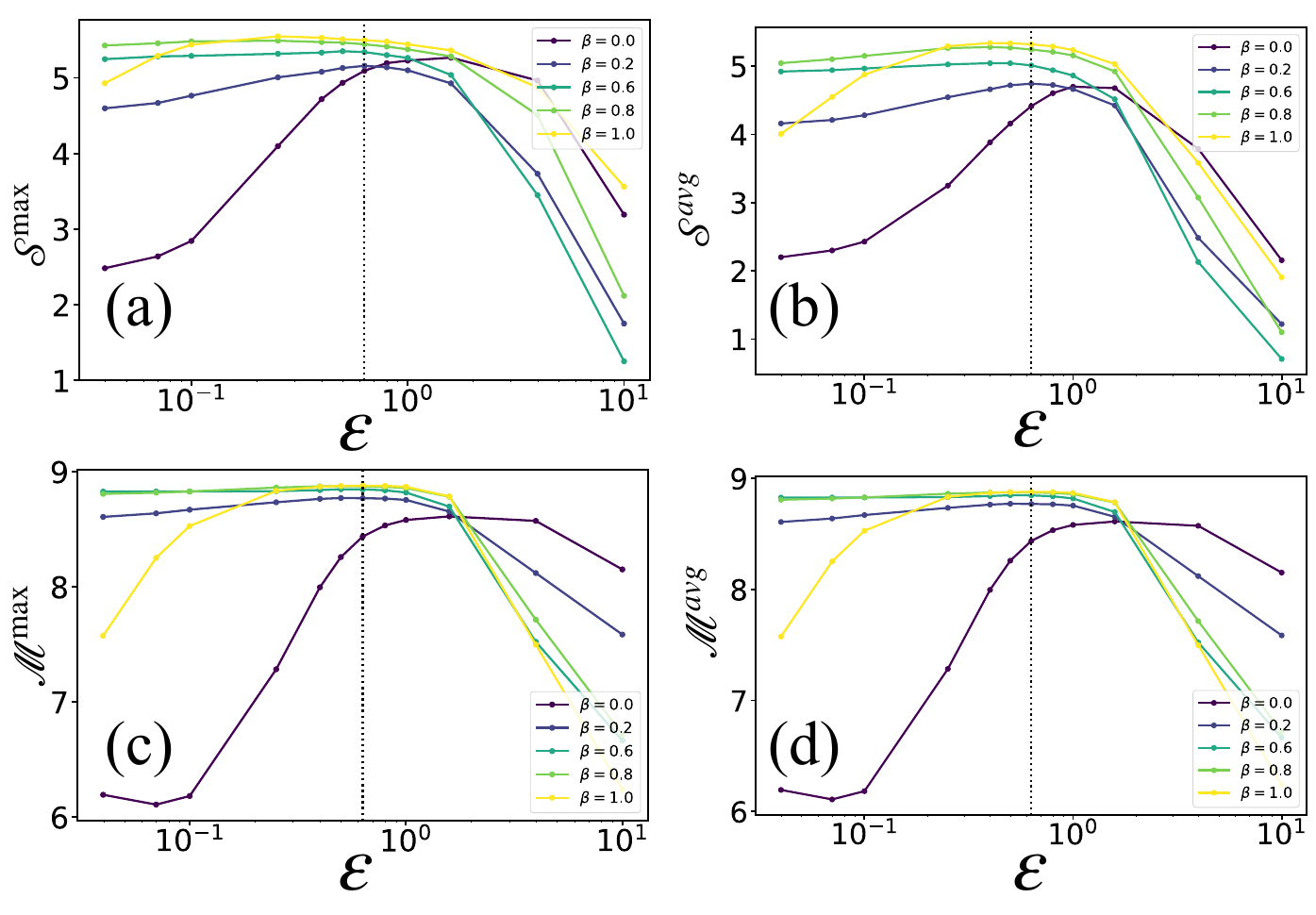}
    \caption{{\bf Maximum and time-averaged quantum resource generation as a function of the impurity strength $\varepsilon$}. (a) and (c) show the maximum bipartite entanglement entropy $\mathcal{S}^{\max}$ and maximum stabilizer R\'enyi entropy $\mathcal{M}^{\max}$ reached during the evolution, respectively, while  (b) and (d) display the corresponding time-averaged values $\mathcal{S}^{\mathrm{avg}}$ and $\mathcal{M}^{\mathrm{avg}}$, respectively. The results are obtained from the dynamics under the pulse-engineered Hamiltonian and averaged over 20 random stabilizer initial states. The comparison between maximum and averaged quantities highlights the distinct response of entanglement and nonstabilizerness to the driving protocol, revealing regimes where their growth is not directly correlated with quantum chaos (compare with the behavior obtained for $\langle r \rangle$ in Fig.~\ref{fig:level_spacing}).
    }
    \label{fig:max_ent_sre}
\end{figure*}

We first examine the entanglement dynamics shown in Figs.~\ref{fig:ent_time}(a)-(d). In the undeformed limit ($\beta=0$), Figs.~\ref{fig:ent_time}(a) and (b) show the evolution of the half-chain entanglement entropy for the same initial state in the low ($\varepsilon=0.03981$)- and high ($\varepsilon=0.63095$)-impurity regimes, respectively. It is clearly seen that the high-impurity regime generates significantly larger entanglement entropy than the low-impurity regime. This enhancement is associated with the onset of quantum chaos at large impurity strength. A very similar trend is observed for the growth of nonstabilizerness as shown in Figs.~\ref{fig:ent_time}(e) and (f). Interestingly, although the ZX model ($\beta=1$) remains non-chaotic for all impurity strengths, Figs.~\ref{fig:ent_time}(c) and (d) show that similar to the  $\mathcal{H}_{\mathrm{XXX}}$ model,  high-impurity regime still helps in faster growth of  entanglement than the low-impurity regime. Though the saturation values at these two impurity strengths remain close to each other. For  nonstabilizerness, in the low-impurity regime ($\varepsilon=0.03981$) (Fig.~\ref{fig:ent_time}(g)), the stabilizer Rényi entropy (SRE) grows relatively slowly. In contrast, in the high-impurity ($\varepsilon=0.63095$,  Fig.~\ref{fig:ent_time}(h)), the SRE saturates rapidly, reaching a value nearly identical to that obtained in the low-impurity case.

In  contrast to the above behavior, similar analysis for pulse-engineered Hamiltonians ($0<\beta<1$) reveals that the dependence of entanglement and nonstabilizerness on spectral chaos becomes much weaker. The behavior of entanglement for two representative intermediate values of $\beta$ is shown in Fig.~\ref{fig:sre_time} ($\beta=0.2$ in (a), (b)) and $\beta=0.8$ in (c), (d)). Unlike the undeformed case ($\beta=0$), where the chaotic regime exhibits noticeably higher entanglement  growth, the entanglement dynamics for low ($\varepsilon=0.03981$)- and high ($\varepsilon=0.63095$)-impurity strengths becomes nearly indistinguishable for intermediate $\beta$, as seen in Figs.~\ref{fig:sre_time}(a)-(d). The stabilizer Rényi entropy displays a similar trend, with both impurity regimes exhibiting almost identical growth and saturation behavior as shown in Figs.~\ref{fig:sre_time}(e)-(h).

 Next, we investigate whether the same behavior persists for a larger ensemble of initial trajectories.  For each trajectory, as described in Eq.~(\ref{eqn:evolution}), we evolve the initial state for a finite time duration. During this evolution, we record the maximum values of the entanglement entropy and the stabilizer Rényi entropy, denoted by $\mathcal{S}^{\max}$ and $\mathcal{M}^{\max}$, respectively. We also compute their time-averaged values $\mathcal{S}^{\mathrm{avg}}$ and $\mathcal{M}^{\mathrm{avg}}$. These quantities provide a quantitative measure of the entanglement- and magic-generation capabilities of the pulse-engineered Hamiltonians. Fig.~\ref{fig:max_ent_sre} summarizes the maximum and time-averaged quantum resources generated during the dynamics for different values of the parameter $\beta$. For the undeformed Hamiltonian ($\beta=0$), both the maximum and average values of entanglement (see Figs.~\ref{fig:max_ent_sre}(a) and (b)) increase sharply as the impurity strength approaches the chaotic regime. As the impurity strength increases further, both quantities decrease noticeably. A similar behavior can be observed for nonstabilizerness depicted in  Figs.~\ref{fig:max_ent_sre}(c) and (d).

A qualitatively different behavior emerges for the pulse-engineered Hamiltonians ($0<\beta<1$). As $\beta$ increases, the maximum and the time-averaged values of both  entanglement entropy and stabilizer R\'enyi entropy become much less sensitive to the impurity strength over a broad parameter range. In particular, for intermediate values of $\beta$, the maximum and time-averaged values remain nearly constant across both the non-chaotic and chaotic regimes. These results demonstrate that pulse engineering largely decouples the generation of quantum resources from the degree of spectral chaos, making the engineered interaction structure the dominant factor governing the dynamics.
%\subsection{Unitary state $k$-design}

\label{sec:k_design}

\section{Discussion}
\label{sec:discussion}

In this work, we investigated how  pulse engineering reshapes the interplay between spectral chaos, entanglement generation, nonstabilizerness, and atypical eigenstates in interacting spin systems. Starting from a chaotic spin-$1/2$ Heisenberg (XXX) chain, we employed average Hamiltonian theory to construct a family of effective Hamiltonians that continuously interpolate between the original XXX model and the ZX Hamiltonian through Clifford pulse sequences.

A central result of this work is the emergence and evolution of structured atypical eigenstates across the engineered Hamiltonians. In the XXX limit, we identified a family of analytically tractable eigenstates forming exact entanglement plateaus with fixed bipartite entanglement entropy, $\mathcal{S}=1$, while exhibiting increasing nonstabilizerness with system size. These states are coherent superpositions of long-range Bell-pair configurations connecting mirror-symmetric sites, resembling resonating-valence-bond-type structures. Despite being embedded in an otherwise chaotic many-body spectrum, they retain highly constrained entanglement patterns and admit exact analytical descriptions. As the pulse strength increases, these exact states continuously evolve into thermal states, while a new set of approximate atypical eigenstates emerges, exhibiting anomalous entanglement  over a broad parameter regime. In the fully pulse-engineered ZX limit, the engineered Hamiltonian provides us with  a pair of exact, long-range-entangled stabilizer eigenstates with vanishing stabilizer Rényi entropy, corresponding to the well-known rainbow scar states.

Beyond revealing these structured eigenstates, our results demonstrate that  pulse engineering provides a powerful means of controlling quantum resource generation independently of spectral chaos. Although the original XXX Hamiltonian exhibits a strong correlation between spectral chaos and the dynamical generation of entanglement and nonstabilizerness, this correlation becomes significantly weaker in the intermediate pulse-engineered regime. Although the level statistics continue to distinguish chaotic and non-chaotic phases, the generated entanglement entropy and stabilizer Rényi entropy remain remarkably similar in both non-chaotic and chaotic regions. These findings reveal a partial decoupling between spectral chaos and quantum resource generation, challenging the widely held expectation that stronger quantum chaos necessarily leads to enhanced quantum resources. 

Therefore, broadly, our work establishes  pulse engineering as a versatile route for generating diverse non-integrable many-body Hamiltonians from a single microscopic model, providing a unified platform for engineering structured eigenstates and controllable quantum resources relevant to emerging quantum technologies. As a future problem, it would  be interesting to investigate the robustness of engineered Hamiltonians against pulse imperfections and environmental noise.

\acknowledgements
We thank Debraj Rakshit for useful discussions. S. S. R. acknowledges financial support from the Faculty Research Scheme, IIT (ISM) Dhanbad, India, under Project No. FRS/2024/PHYSICS/MISC0110, and from the Anusandhan National Research Foundation (ANRF), Government of India, under Grant Nos. ANRF/ARG/2025/004617/PS and ANRF/ECRG/2025/002793/PMS. We also acknowledge the use of the QuSpin package~\cite{QuSpin} for the numerical simulations presented  in this work.

\appendix

\section{Derivation of the Effective Hamiltonian}
\label{sec:model_Derivation}
In this appendix, we derive the target  Hamiltonian $\mathcal{H}_{\mathrm{Target}}$ from the actual evolution exploiting the Magnus expansion.  The undeformed  Hamiltonian we start with is given by 
\begin{equation}
\mathcal{H}_{\mathrm{XXX}} =
\sum_{i=1}^{L-1}
J\Big(\sigma^x_i \sigma^x_{i+1}+\sigma^y_i \sigma^y_{i+1}
+
\sigma^z_i \sigma^z_{i+1}
\Big)
+
\varepsilon \sigma^z_m .
\end{equation}

The evolution operator over one driving period is given by
Eq.~\eqref{eqn:evolution_exact} of the main text. Throughout this
appendix, we employ the leading-order average Hamiltonian,
\begin{equation}
\mathcal{H}_{\mathrm{Target}}
=
\mathcal{H}_{\mathrm{eff}}^{(0)}
=
\frac{1}{T}
\sum_k
\mathcal{H}_k t_k,
\qquad
\mathcal{H}_k=g_k^\dagger\mathcal{H}g_k,
\label{eq:Htarget_appendix}
\end{equation}
where $T=\sum_k t_k$ is the duration of one driving cycle. As
demonstrated in Appendix~\ref{app:floquet_match}, the higher-order
Magnus corrections are negligible for the driving periods considered in
this work. We therefore derive the explicit form of
$\mathcal{H}_{\mathrm{Target}}$ for the pulse sequences used below.

As in our protocol, we consider only an odd number of lattice sites $L$, two distinct cases arise depending on whether the middle site $m=(L+1)/2$ is even or odd. Here, we only show the derivation when $m$ is even.  In this case, the pulse operators $p_k$ and the corresponding $g_k$-operators are given explicitly in Eqs.~\eqref{eq:pulses1} and \eqref{eq:g1}, of the main text respectively.

\noindent The Hadamard rotation on the Pauli operators acts as follows: 
\begin{eqnarray}  H_i\sigma^x_i H_i = \sigma^z_i, H_i\sigma^z_iH_i = \sigma^x_i, 
 H_i\sigma^y_iH_i=-\sigma^y_i. \end{eqnarray}  As a result, in the first interval $t_0$, when $g_0 = I_1\otimes H_2\otimes I_3\otimes H_4\otimes\cdots$ acts on the original Hamiltonian, it transforms to 

%the interaction terms in the XXX Hamiltonian transforms as 
%\begin{eqnarray}
%\sigma^x_i  \sigma^x_{i+1}&\rightarrow & \sigma^z_i  \sigma^x_{i+1}, \nonumber\\
%\sigma^z_i  \sigma^z_{i+1} &\rightarrow & \sigma^x_i  \sigma^z_{i+1}, \nonumber\\
%\sigma^y_i  \sigma^y_{i+1} &\rightarrow & -\sigma^y_i  %\sigma^y_{i+1}.
%\end{eqnarray}
\begin{eqnarray}
\mathcal{H}_{\mathrm{0}} = g_{0}^\dagger\, \mathcal{H}_{\mathrm{XXX}}\, g_{0}&=&
\sum_{i=1}^{L-1}
J\left(\sigma^z_i \sigma^x_{i+1}-\sigma^y_i\sigma^y_{i+1}
+\sigma^x_i \sigma^z_{i+1}\right) \nonumber \\
&+&\varepsilon \sigma^x_m .
\end{eqnarray}

\noindent Similarly,  under the action of $\sigma^y$, the Pauli matrices transform as, 
\begin{eqnarray}  \sigma^y_i\sigma^x_i\sigma^y_i =- \sigma^x_i, \sigma^y_i\sigma^y_i\sigma^y_i = \sigma^y_i,
  \sigma^y_i\sigma^z_i\sigma^y_i=-\sigma^z_i. \end{eqnarray} Hence, in the second interval $t_1$, when $g_1 = I_1\otimes(\sigma^{y})_2\otimes I_3\otimes(\sigma^{y})_4\otimes\cdots$  acts  on the original Hamiltonian, we get 

\begin{eqnarray}
\mathcal{H}_{\mathrm{1}} = g_{1}^\dagger\, \mathcal{H}_{\mathrm{XXX}}\, g_{1}&=&
\sum_{i=1}^{L-1}
J\left(-\sigma^x_i \sigma^x_{i+1}
+\sigma^y_i \sigma^y_{i+1}-\sigma^z_i\sigma^z_{i+1}\right)
\nonumber \\&-&\varepsilon \sigma^z_m .
\end{eqnarray}

In the last interval $t_2$, $g_2$ is the identity operator on all sites and the actions are trivial, leading to 
\begin{eqnarray}
\mathcal{H}_{\mathrm{2}} = g_{2}^\dagger\,\mathcal{H}_{\mathrm{XXX}}\, g_{2} &=&
\sum_{i=1}^{L-1}
J\left(\sigma^x_i \sigma^x_{i+1}+\sigma^y_i \sigma^y_{i+1}+\sigma^z_i \sigma^z_{i+1}
\right)
\nonumber \\ &+&\varepsilon \sigma^z_m .
\end{eqnarray}

The effective Hamiltonian is obtained from the lowest-order average Hamiltonian expansion,
\begin{equation}
\mathcal{H}_{\mathrm{Target}}
=
\frac{t_0 \mathcal{H}_0+t_1 \mathcal{H}_1+t_2 \mathcal{H}_2}{T},
\end{equation}
where $T=t_0+t_1+t_2$ is the one  pulse  cycle duration.

Substituting the explicit forms of $\mathcal{H}_0$, $\mathcal{H}_1$, and $\mathcal{H}_2$ into the average Hamiltonian expression,
we obtain
\begin{eqnarray}
\mathcal{H}_{\mathrm{Target}}
&=&\sum_{i=1}^{L-1}
\Bigg[
\frac{J(t_2-t_1)}{T}\,
\sigma^x_i\sigma^x_{i+1} +
\frac{J(t_1+t_2-t_0)}{T}\,
\sigma^y_i\sigma^y_{i+1}\nonumber \\&+&
\frac{J(t_2-t_1)}{T}\,
\sigma^z_i\sigma^z_{i+1}
+ \frac{Jt_0}{T}\,
\sigma^z_i\sigma^x_{i+1}
+
\frac{Jt_0}{T}\,
\sigma^x_i\sigma^z_{i+1} \nonumber \\&+&
\frac{\varepsilon(t_2-t_1)}{T}\sigma^z_m
+
\frac{\varepsilon t_0}{T}\sigma^x_m \Bigg].
\end{eqnarray}
Let us now choose $t_0=2t_1$, and the above expression becomes 
\begin{eqnarray}
\mathcal{H}_{\mathrm{Target}}
&=&\sum_{i=1}^{L-1}\Bigg[
\frac{(t_2-t_1)}{T}(
J\sigma^x_i\sigma^x_{i+1}+
J\sigma^y_i\sigma^y_{i+1}\nonumber\\ &+&
J\sigma^z_i\sigma^z_{i+1}+\sigma^z_m)
+\frac{2t_1}{T}
(J\sigma^z_i\sigma^x_{i+1}
+J\sigma^x_i\sigma^z_{i+1} \nonumber \\ &+&
\sigma^x_m )\Bigg].
\end{eqnarray}

\noindent Now, denoting $t_1/t_2=\beta$, we get 
\begin{align}
\mathcal{H}_{\mathrm{Target}}
&=\frac{t_2}{T}
\Bigg[
(1-\beta) \Bigg(\sum_{i=1}^{L-1}(J
\sigma^x_i\sigma^x_{i+1}+\sigma^y_i\sigma^y_{i+1}+
\sigma^z_i\sigma^z_{i+1})\nonumber \\ &+ \varepsilon\sigma^z_m\Bigg)
+
2\beta \Bigg(\sum_{i=1}^{L-1}J(\sigma^z_i\sigma^x_{i+1}
+J\sigma^x_i\sigma^z_{i+1})+\varepsilon \sigma^x_m\Bigg)\Bigg].
\label{eq:Heff_full}
\end{align}

\noindent Excluding this global factor $t_2/T$, we finally get 
\begin{equation}
\mathcal{H}_{\mathrm{Target}}
=(1-2\beta)\mathcal{H}_{\mathrm{XXX}}
+2\beta  \mathcal{H}_{\mathrm{ZX}},
\label{eq:H_eff}
\end{equation}
where 
\begin{equation}
\mathcal{H}_{\mathrm{ZX}} =
\sum_{i=1}^{L-1}
J\left(
\sigma^z_i \sigma^x_{i+1}
+
\sigma^x_i \sigma^z_{i+1}
\right)
+
\varepsilon \sigma^x_m. 
\label{eq:H2_general_appendix}
\end{equation}

Applying the same procedure, we can get the effective Hamiltonian which is same as in Eq.~\eqref{eq:H_eff}, when $m$ is odd. Hence, Eq.~\eqref{eq:H_eff} holds for all odd values of $L$.

\section{Validity of the Leading-Order Magnus Approximation}
\label{app:floquet_match}

In this appendix, we justify the use of the leading-order average Hamiltonian employed throughout the main text. The exact evolution operator over one driving period $T$, given in Eq.~(\ref{eqn:evolution_exact}), can be written compactly as
\begin{equation}
U_T
=
g_n
e^{-i\mathcal{H}_nt_n}
e^{-i\mathcal{H}_{n-1}t_{n-1}}
\cdots
e^{-i\mathcal{H}_1t_1}
e^{-i\mathcal{H}_0t_0},
\label{eq:UT_compact}
\end{equation}
where $g_n=p_np_{n-1}\cdots p_0$ and
$\mathcal{H}_k=g_k^\dagger \mathcal{H} g_k$ is the Hamiltonian in the corresponding toggling frame.

According to average Hamiltonian theory, the  evolution over one driving cycle is described by
\begin{equation}
U_T=e^{-i\mathcal{H}_{\mathrm{eff}}T},
\end{equation}
where the effective Hamiltonian is given by the Magnus expansion
\begin{align}
\mathcal{H}_{\mathrm{eff}}
&=
\mathcal{H}_{\mathrm{eff}}^{(0)}
+
\mathcal{H}_{\mathrm{eff}}^{(1)}
+
\mathcal{H}_{\mathrm{eff}}^{(2)}
+\cdots
\nonumber\\
&=
\frac{1}{T}\sum_{k} \mathcal{H}_k t_k
-\frac{i}{2T}
\sum_{k>l}
[\mathcal{H}_k,\mathcal{H}_l]t_kt_l
+\mathcal{O}(T^2),
\label{eq:Magnus_appendix}
\end{align}
with the leading-order term
\begin{equation}
\mathcal{H}_{\mathrm{Target}}=\mathcal{H}_{\mathrm{eff}}^{(0)}
=\frac{1}{T}\sum_{k}\mathcal{H}_k t_k .
\end{equation}

Therefore, in our case, the target Hamiltonian is identified with the zeroth-order Magnus Hamiltonian,
\begin{equation}
\mathcal{H}_{\mathrm{Target}}
\equiv
\mathcal{H}_{\mathrm{eff}}^{(0)}.
\end{equation}
The explicit construction of $H_{\mathrm{Target}}$ for the pulse sequences employed in this work is presented in Appendix~\ref{sec:model_Derivation}.\\

\begin{figure}
    \centering
    \includegraphics[width=\linewidth]{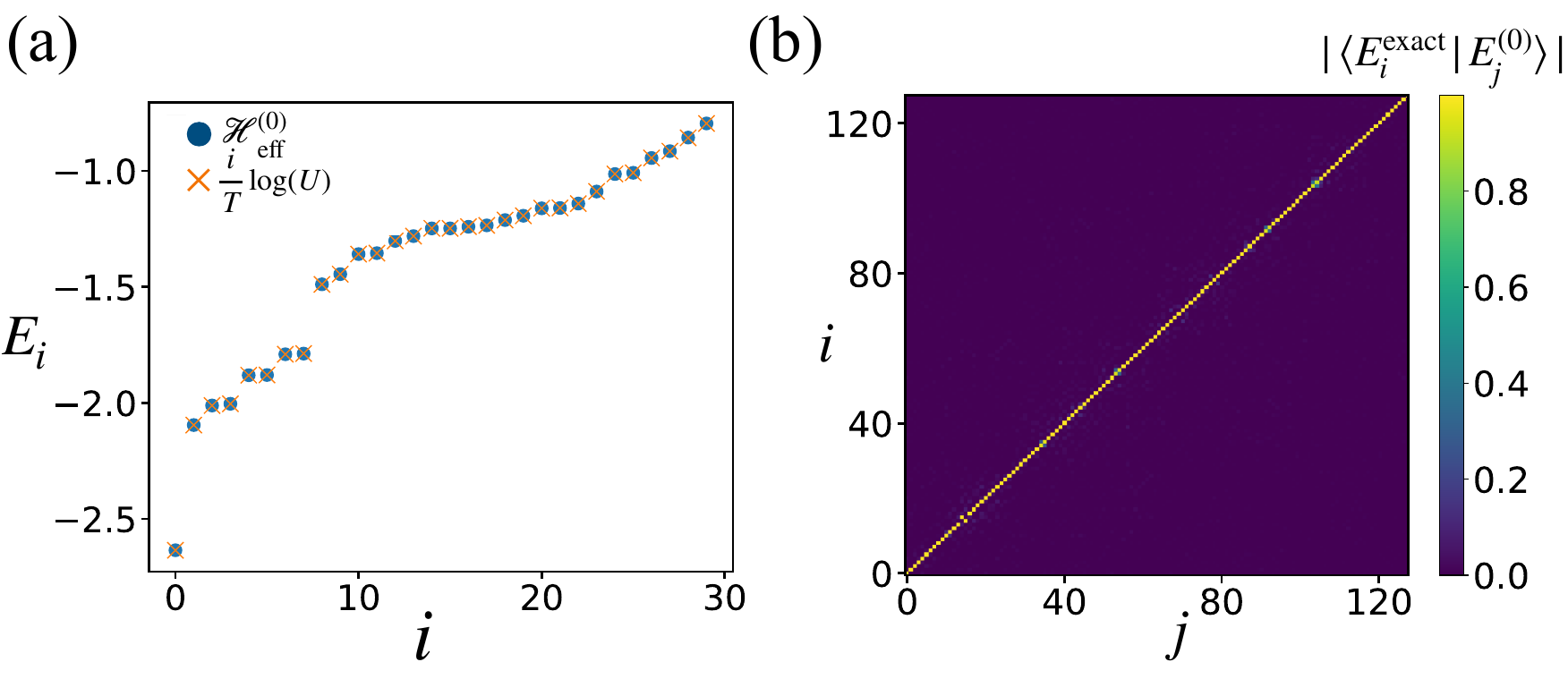}
    \caption{\textbf{Validation of the leading-order Magnus approximation.}
    (a) Comparison between the eigenvalues of the exact  Hamiltonian, obtained from the evolution operator $U_T$ (Eq.~(\ref{eqn:H_eff})), and those of the zeroth-order average Hamiltonian $\mathcal{H}_{\mathrm{eff}}^{(0)}$ for $T=0.1$, $L=7$, $\varepsilon=0$, and  $\beta=0.4$. The excellent agreement demonstrates that the quasienergy spectrum is accurately captured by the leading-order effective Hamiltonian.
    (b) Overlap matrix $|\langle E_i^{\mathrm{exact}}|E_j^{(0)}\rangle|$ between the eigenstates of the exact  Hamiltonian and $\mathcal{H}_{\mathrm{eff}}^{(0)}$. The nearly perfect diagonal structure indicates a one-to-one correspondence between the eigenstates of the two Hamiltonians, confirming that the leading-order Magnus approximation accurately describes the  dynamics in the small-$T$ regime.}
    \label{fig:magnus}
\end{figure}

Figure~\ref{fig:magnus} validates the leading-order approximation used throughout this work. (a) Compares the spectrum of the Hamiltonian,  $\mathcal{H}_{\mathrm{eff}}=\frac{i}{T}\log(U_T)$  with that of $\mathcal{H}_{\mathrm{eff}}^{(0)}$, while  (b) compares the corresponding eigenstates through their overlap matrix, $|\langle E_i^{\mathrm{exact}}|E_j^{(0)}\rangle|$. The excellent agreement in both the eigenvalues and eigenvectors demonstrates that higher-order Magnus corrections are negligible for the driving period considered here. Consequently, the  dynamics are accurately captured by the target Hamiltonian. Thereby justifying the leading-order approximation used in the main text.\\

\newpage

\begin{widetext}
\section{Entanglement and nonstabilizerness plateau for higher system size}
\label{app:platau}

We would like to remind the reader that although the exact form of the $L-1$ states in the XXX limit can be derived from the effective Hamiltonian description, for better  understanding we  present their explicit form for a particular higher system size $L=13$.
For this system size, the same entanglement plateau persists while the states develop increasingly intricate long-range Bell-pair superpositions.  %In this case, twelve such eigenstates appear:

In this case, twelve such eigenstates appear, with $\mathcal{M}=3.62154$, constructed from superpositions of six long-range Bell-pair coverings:

\label{L=13 states}
\begin{align}
|\psi_{1}\rangle
&= \ket{1}_7 \Bigl(
a\ket{\Psi^-}_{1,13}\ket{1}_{2345689(10)(11)(12)}
- b\ket{\Psi^-}_{2,12}\ket{1}_{1345689(10)(11)(13)}
+ c\ket{\Psi^-}_{3,11}\ket{1}_{1245689(10)(12)(13)}
\nonumber\\
&\hspace{2.3cm}
- d\ket{\Psi^-}_{4,10}\ket{1}_{1235689(11)(12)(13)}
+ e\ket{\Psi^-}_{5,9}\ket{1}_{123468(10)(11)(12)(13)}
- f\ket{\Psi^-}_{6,8}\ket{1}_{123459(10)(11)(12)(13)}
\Bigr), \nonumber\\
|\psi_{2}\rangle
&= \ket{0}_7 \Bigl(
a\ket{\Psi^-}_{1,13}\ket{0}_{2345689(10)(11)(12)}
- b\ket{\Psi^-}_{2,12}\ket{0}_{1345689(10)(11)(13)}
+ c\ket{\Psi^-}_{3,11}\ket{0}_{1245689(10)(12)(13)}
\nonumber\\
&\hspace{2.3cm}
- d\ket{\Psi^-}_{4,10}\ket{0}_{1235689(11)(12)(13)}
+ e\ket{\Psi^-}_{5,9}\ket{0}_{123468(10)(11)(12)(13)}
- f\ket{\Psi^-}_{6,8}\ket{0}_{123459(10)(11)(12)(13)}
\Bigr), \nonumber\\
|\psi_{3}\rangle
&= \ket{1}_7 \Bigl(
f\ket{\Psi^-}_{1,13}\ket{1}_{2345689(10)(11)(12)}
- d\ket{\Psi^-}_{2,12}\ket{1}_{1345689(10)(11)(13)}
+ b\ket{\Psi^-}_{3,11}\ket{1}_{1245689(10)(12)(13)}
\nonumber\\
&\hspace{2.3cm}
+ a\ket{\Psi^-}_{4,10}\ket{1}_{1235689(11)(12)(13)}
- c\ket{\Psi^-}_{5,9}\ket{1}_{123468(10)(11)(12)(13)}
+ e\ket{\Psi^-}_{6,8}\ket{1}_{123459(10)(11)(12)(13)}
\Bigr), \nonumber\\
|\psi_{4}\rangle
&= \ket{0}_7 \Bigl(
f\ket{\Psi^-}_{1,13}\ket{0}_{2345689(10)(11)(12)}
- d\ket{\Psi^-}_{2,12}\ket{0}_{1345689(10)(11)(13)}
+ b\ket{\Psi^-}_{3,11}\ket{0}_{1245689(10)(12)(13)}
\nonumber\\
&\hspace{2.3cm}
+ a\ket{\Psi^-}_{4,10}\ket{0}_{1235689(11)(12)(13)}
- c\ket{\Psi^-}_{5,9}\ket{0}_{123468(10)(11)(12)(13)}
+ e\ket{\Psi^-}_{6,8}\ket{0}_{123459(10)(11)(12)(13)}
\Bigr), \nonumber\\
|\psi_{5}\rangle
&= \ket{1}_7 \Bigl(
+ b\ket{\Psi^-}_{1,13}\ket{1}_{2345689(10)(11)(12)}
- e\ket{\Psi^-}_{2,12}\ket{1}_{1345689(10)(11)(13)}
- f\ket{\Psi^-}_{3,11}\ket{1}_{1245689(10)(12)(13)}
\nonumber\\
&\hspace{2.3cm}
+ c\ket{\Psi^-}_{4,10}\ket{1}_{1235689(11)(12)(13)}
+ a\ket{\Psi^-}_{5,9}\ket{1}_{123468(10)(11)(12)(13)}
- d\ket{\Psi^-}_{6,8}\ket{1}_{123459(10)(11)(12)(13)}
\Bigr), \nonumber\\
|\psi_{6}\rangle
&= \ket{0}_7 \Bigl(
+ b\ket{\Psi^-}_{1,13}\ket{0}_{2345689(10)(11)(12)}
- e\ket{\Psi^-}_{2,12}\ket{0}_{1345689(10)(11)(13)}
- f\ket{\Psi^-}_{3,11}\ket{0}_{1245689(10)(12)(13)}
\nonumber\\
&\hspace{2.3cm}
+ c\ket{\Psi^-}_{4,10}\ket{0}_{1235689(11)(12)(13)}
+ a\ket{\Psi^-}_{5,9}\ket{0}_{123468(10)(11)(12)(13)}
- d\ket{\Psi^-}_{6,8}\ket{0}_{123459(10)(11)(12)(13)}
\Bigr), \nonumber\\
|\psi_{7}\rangle
&= \ket{1}_7 \Bigl(
e\ket{\Psi^-}_{1,13}\ket{1}_{2345689(10)(11)(12)}
- a\ket{\Psi^-}_{2,12}\ket{1}_{1345689(10)(11)(13)}
- d\ket{\Psi^-}_{3,11}\ket{1}_{1245689(10)(12)(13)}
\nonumber\\
&\hspace{2.3cm}
- f\ket{\Psi^-}_{4,10}\ket{1}_{1235689(11)(12)(13)}
+ b\ket{\Psi^-}_{5,9}\ket{1}_{123468(10)(11)(12)(13)}
+ c\ket{\Psi^-}_{6,8}\ket{1}_{123459(10)(11)(12)(13)}
\Bigr), \nonumber\\
|\psi_{8}\rangle
&= \ket{0}_7 \Bigl(
e\ket{\Psi^-}_{1,13}\ket{0}_{2345689(10)(11)(12)}
- a\ket{\Psi^-}_{2,12}\ket{0}_{1345689(10)(11)(13)}
- d\ket{\Psi^-}_{3,11}\ket{0}_{1245689(10)(12)(13)}
\nonumber\\
&\hspace{2.3cm}
- f\ket{\Psi^-}_{4,10}\ket{0}_{1235689(11)(12)(13)}
+ b\ket{\Psi^-}_{5,9}\ket{0}_{123468(10)(11)(12)(13)}
+ c\ket{\Psi^-}_{6,8}\ket{0}_{123459(10)(11)(12)(13)}
\Bigr), \nonumber\\
|\psi_{9}\rangle
&= \ket{1}_7 \Bigl(
c\ket{\Psi^-}_{1,13}\ket{1}_{2345689(10)(11)(12)}
+ f\ket{\Psi^-}_{2,12}\ket{1}_{1345689(10)(11)(13)}
- a\ket{\Psi^-}_{3,11}\ket{1}_{1245689(10)(12)(13)}
\nonumber\\
&\hspace{2.3cm}
- e\ket{\Psi^-}_{4,10}\ket{1}_{1235689(11)(12)(13)}
- d\ket{\Psi^-}_{5,9}\ket{1}_{123468(10)(11)(12)(13)}
- b\ket{\Psi^-}_{6,8}\ket{1}_{123459(10)(11)(12)(13)}
\Bigr), \nonumber\\
|\psi_{10}\rangle
&= \ket{0}_7 \Bigl(
c\ket{\Psi^-}_{1,13}\ket{0}_{2345689(10)(11)(12)}
+ f\ket{\Psi^-}_{2,12}\ket{0}_{1345689(10)(11)(13)}
- a\ket{\Psi^-}_{3,11}\ket{0}_{1245689(10)(12)(13)}
\nonumber\\
&\hspace{2.3cm}
- e\ket{\Psi^-}_{4,10}\ket{0}_{1235689(11)(12)(13)}
- d\ket{\Psi^-}_{5,9}\ket{0}_{123468(10)(11)(12)(13)}
- b\ket{\Psi^-}_{6,8}\ket{0}_{123459(10)(11)(12)(13)}
\Bigr), \nonumber\\
|\psi_{11}\rangle
&= \ket{1}_7 \Bigl(
d\ket{\Psi^-}_{1,13}\ket{1}_{2345689(10)(11)(12)}
+ c\ket{\Psi^-}_{2,12}\ket{1}_{1345689(10)(11)(13)}
+ e\ket{\Psi^-}_{3,11}\ket{1}_{1245689(10)(12)(13)}
\nonumber\\
&\hspace{2.3cm}
+ b\ket{\Psi^-}_{4,10}\ket{1}_{1235689(11)(12)(13)}
+ f\ket{\Psi^-}_{5,9}\ket{1}_{123468(10)(11)(12)(13)}
+ a\ket{\Psi^-}_{6,8}\ket{1}_{123459(10)(11)(12)(13)}
\Bigr), \nonumber\\
|\psi_{12}\rangle
&= \ket{0}_7 \Bigl(
d\ket{\Psi^-}_{1,13}\ket{0}_{2345689(10)(11)(12)}
+ c\ket{\Psi^-}_{2,12}\ket{0}_{1345689(10)(11)(13)}
+ e\ket{\Psi^-}_{3,11}\ket{0}_{1245689(10)(12)(13)}
\nonumber\\
&\hspace{2.3cm}
+ b\ket{\Psi^-}_{4,10}\ket{0}_{1235689(11)(12)(13)}
+ f\ket{\Psi^-}_{5,9}\ket{0}_{123468(10)(11)(12)(13)}
+ a\ket{\Psi^-}_{6,8}\ket{0}_{123459(10)(11)(12)(13)}
\Bigr).
\end{align}

The coefficients are
\begin{equation}
a=0.13275,\quad
b=0.36784,\quad
c=0.51865,\quad
d=0.55066,\quad
e=0.45651,\quad
f=0.25778.
\end{equation}

\end{widetext}

\section{Clifford Equivalence of the Special States}
\label{app:clifford}

The special eigenstates of XXX model discussed in the main text were found to possess same stabilizer R\'enyi entropy. This observation suggests that these states are likely related by Clifford transformations. In this appendix, we explicitly construct such transformations for several representative cases.

The transformations used throughout this appendix consist only of SWAP gates, products of local Pauli operators, and the global spin-flip operator
\begin{equation}
X_g=\prod_{i=1}^{L}\sigma_i^x .
\end{equation}

\noindent Since all of these operations belong to the Clifford group, the mappings established below demonstrate the Clifford equivalence of the corresponding states.

%\subsection{$L=3$}
\subsection{\texorpdfstring{$L=3$}{L=3}}

\noindent The two special states are related simply by a global spin flip,
\begin{equation}
|\psi_{1}\rangle=X_g|\psi_{2}\rangle.
\end{equation}

%\subsection{$L=5$}
\subsection{\texorpdfstring{$L=5$}{L=5}}

\noindent For $L=5$, consider
\begin{equation}
|\psi_{1}\rangle=
|1\rangle_3
\left(
a|\Psi^{-}\rangle_{15}|11\rangle_{24}
-
b|\Psi^{-}\rangle_{24}|11\rangle_{15}
\right).
\end{equation}

\noindent Applying the Clifford operation
\begin{equation}
\mathcal{U}=
\sigma_z^1 \sigma_z^5
(\mathrm{SWAP})_{1,2}
(\mathrm{SWAP})_{4,5},
\end{equation}
one obtains
\begin{equation}
\mathcal{U}|\psi_{1}\rangle
=
|\psi_{3}\rangle.
\end{equation}

\noindent In addition,
\begin{equation}
|\psi_{2}\rangle=X_g|\psi_{1}\rangle\ \text{and}  \\ 
\ |\psi_{4}\rangle=X_g|\psi_{3}\rangle.
\end{equation}

\noindent Hence, all four states belong to the same set of Clifford states.

%\subsection{$L=7$}

%\begin{equation}
%|\psi_{56}\rangle=|1\rangle_4\left(a|\Psi^{}\rangle_{17}|1\rangle_{2356}-%b|\Psi^{-%}\rangle_{26}|1\rangle_{1357}+c|\Psi^{}\rangle_{35}|1\rangle_{1267}\right)
%\end{equation}
%is mapped to $|\psi_{114}\rangle$ through

%\begin{equation}
%U=\sigma_z^3\sigma_z^5({\rm SWAP})_{2,3}({\rm SWAP})_{5,6}({\rm SWAP})_{1,2}
%({\rm SWAP})_{6,7},
%\end{equation}
%such that
%\begin{equation}
%U|\psi_{56}\rangle=|\psi_{114}\rangle.
%\end{equation}

%Furthermore,
%\begin{equation}
%|\psi_{61}\rangle=X_g|\psi_{56}\rangle .
%\end{equation}

%Therefore these states are Clifford equivalent.

%\subsection{$L=9$}

%For $L=9$, the state $|\psi_{288}\rangle$ is transformed into
%|\psi_{356}\rangle$ by

%\begin{equation}
%U=\sigma_z^1 \sigma_z^3({\rm SWAP})_{1,4}({\rm SWAP})_{6,9}
%({\rm SWAP})_{1,3}({\rm SWAP})_{7,9},
%\end{equation}
%giving
%\begin{equation}
%U|\psi_{288}\rangle=|\psi_{356}\rangle .
%\end{equation}

%The global spin-flip operation generates another member of the family,
%\begin{equation}
%|\psi_{295}\rangle=X_g|\psi_{288}\rangle .
%\end{equation}

%\subsection{$L=11$}

%Similarly, for $L=11$, the state $|\psi_{1354}\rangle$ is mapped to
%$|\psi_{1553}\rangle$ by a sequence of SWAP operations followed by a local Pauli correction,
%\begin{equation}
%|\psi_{1553}\rangle=
%\sigma_z^2 \sigma_z^4\,U_{\rm swap}
%|\psi_{1354}\rangle ,
%\end{equation}
%where $U_{\rm SWAP}$ denotes the corresponding product of SWAP gates given in the text.

%% \begin{equation}
%|\psi_{1385}\rangle
%=
%X_g|\psi_{1354}\rangle .
%\end{equation}

%\subsection{$L=13$}
\subsection{\texorpdfstring{$L=13$}{L=13}}
\noindent For $L=13$, the state $|\psi_{1}\rangle$ is mapped to
$|\psi_{3}\rangle$ through the following Clifford operation 

\begin{equation}
|\psi_{3}\rangle
=\sigma_z^1 \sigma_z^3 \sigma_z^5 \sigma_z^9\sigma_z^{11}\sigma_z^{13}
\mathcal{U}_{\mathrm SWAP}
|\psi_{1}\rangle ,
\end{equation}
where $\mathcal{U}_{\mathrm SWAP}$ denotes the sequence of SWAP gates i.e.

\begin{align}
\mathcal{U}_{\mathrm{SWAP}}
=
\prod_{(i,j)\in\mathcal{R}}
\mathrm{SWAP}_{i,j},
%\qquad
%\mathcal{R}
%= \big\{&(1,6),(8,13),(2,4),\nonumber\\
%&(10,12),(3,4),(10,11),\nonumber\\
%&(5,6),(8,9),(4,5),(9,10)
%\big\}.
\end{align}

\noindent Here, $\mathcal{R} =$ $\big\{(1,6),$ $(8,13),$ $(2,4),$ $(10,12),$ $(3,4),$ $(10,11),$ $(5,6),$ $(8,9),$ $(4,5),$ $(9,10)\big\}$ contains all the indices of SWAP gates to be acted sequentially ($\mathrm{SWAP}_{1,6}$ acts first and $\mathrm{SWAP}_{9,10}$ acts last).

\noindent In addition,
\begin{equation}
|\psi_{2}\rangle
=X_g|\psi_{1}\rangle .
\end{equation}
In the same manner, we can map all the other atypical  states  obtained for L=13 following certain  Clifford operations.
\bibliography{references}{}

\end{document}